\documentclass[journal]{IEEEtran}

\usepackage[T1]{fontenc}
\usepackage{textcomp}
\usepackage[english]{babel}
\usepackage{microtype}

\usepackage{amssymb,mathtools,bm}
\usepackage{siunitx}

\usepackage{booktabs}
\usepackage{multirow}
\usepackage{array}
\usepackage[table]{xcolor}
\usepackage{tabularx}
\usepackage{float}

\usepackage{graphicx}
\usepackage{caption}
\usepackage{subcaption}
\usepackage{enumitem}

\usepackage{algorithm}
\usepackage{algpseudocode}
\usepackage{listings}

\usepackage{cite}
\usepackage[colorlinks,citecolor=blue,urlcolor=blue,linkcolor=blue]{hyperref}

\begin{document}

\title{Performance Analysis and Deployment Considerations of Post-Quantum Cryptography for Consumer Electronics}

\author{Daniel Commey,
        Benjamin Appiah,
        Griffith S. Klogo,
        Winful Bagyl-Bac,
        James D. Gadze,
        Yousef Alsenani,
        and~Garth V. Crosby%

\thanks{D. Commey is with the Department of Multidisciplinary Engineering, Texas A\&M University, College Station, TX, USA (e-mail: \texttt{dcommey@tamu.edu}).}%
\thanks{B. Appiah is with the Department of Computer Science, Ho Technical University, Ho, Volta Region, Ghana (e-mail: \texttt{bappiah@htu.edu.gh}).}%
\thanks{G. S. Klogo is with the Department of Computer Engineering, and J. D. Gadze is with the Department of Telecommunication Engineering, Kwame Nkrumah University of Science and Technology (KNUST), Kumasi, Ghana (e-mail: \texttt{gsklogo.coe@knust.edu.gh}; \texttt{jdgadze.coe@knust.edu.gh}).}%
\thanks{W. Bagyl-Bac is with the Department of Computer Science, George Washington University, Washington, DC, USA (e-mail: \texttt{winful.bagylbac@gwu.edu}).}%
\thanks{Y. Alsenani is with the Department of Information Systems, Faculty of Computing and Information Technology, King Abdulaziz University, Jeddah, Saudi Arabia, and also with EleeN LLC, Riyadh, Saudi Arabia (e-mail: \texttt{yalsenani@kau.edu.sa}).}%
\thanks{G. V. Crosby is with the Department of Engineering Technology \& Industrial Distribution, Texas A\&M University, College Station, TX, USA (e-mail: \texttt{gvcrosby@tamu.edu}).}%
}

\maketitle

\begin{abstract}
Quantum computing threatens the security foundations of consumer electronics (CE). Preparing the diverse CE ecosystem, particularly resource-constrained devices, for the post-quantum era requires quantitative understanding of quantum-resistant cryptography (PQC) performance. This paper presents a comprehensive cross-platform performance analysis of leading PQC Key Encapsulation Mechanisms (KEMs) and digital signatures (NIST standards/candidates) compared against classical RSA/ECC. We evaluated execution time, communication costs (key/signature sizes), and memory footprint indicators on high-performance (macOS/M4, Ubuntu/x86) and constrained platforms (Raspberry Pi 4/ARM). Our quantitative results reveal lattice-based schemes, notably NIST standards ML-KEM (Kyber) and ML-DSA (Dilithium), provide a strong balance of computational efficiency and moderate communication/storage overhead, making them highly suitable for many CE applications. In contrast, code-based Classic McEliece imposes significant key size challenges, while hash-based SPHINCS+ offers high security assurance but demands large signature sizes impacting bandwidth and storage. Based on empirical data across platforms and security levels, we provide specific deployment recommendations tailored to different CE scenarios (e.g., wearables, smart home hubs, mobile devices), offering guidance for manufacturers navigating the PQC transition.
\end{abstract}


\section{Introduction}
\IEEEPARstart{T}{he} proliferation of connected consumer electronics (CE) devices—smartphones, wearables, smart home assistants, connected vehicles, and countless Internet of Things (IoT) gadgets—has revolutionized daily life but also expanded the attack surface for security threats \cite{keranen_cryptographic_2014,bui_health_2011}. The security of these devices heavily relies on classical public-key cryptography, primarily RSA and Elliptic Curve Cryptography (ECC), for tasks like secure communication, software updates, authentication, and data protection \cite{stallings_cryptography_2017}. However, the advent of practical quantum computers poses an existential threat to these cryptographic foundations. Shor's algorithm, executable on a sufficiently powerful quantum computer, can efficiently break RSA and ECC, rendering vast amounts of currently secured CE data and communication channels vulnerable \cite{shor_polynomial-time_1999}.

While the exact timeline for fault-tolerant quantum computers remains debated \cite{mosca_quantum_2021}, the "harvest now, decrypt later" threat necessitates immediate action. Sensitive data transmitted or stored by CE devices today could be captured and decrypted retrospectively once quantum computers become available \cite{grumbling_quantum_2019}. Furthermore, the long lifecycle of many CE products means devices sold today must remain secure against future quantum threats. This urgency has driven the development and standardization of post-quantum cryptography (PQC)—algorithms designed to resist attacks from both classical and quantum computers \cite{bernstein_post-quantum_2017}. The U.S. National Institute of Standards and Technology (NIST) has been leading a multi-year effort to select and standardize PQC algorithms, recently announcing initial standards \cite{alagic_status_2022}, which have culminated in official publications for schemes like the Module-Lattice-Based Digital Signature Standard (ML-DSA) \cite{NIST_FIPS_204} and the Stateless Hash-Based Digital Signature Standard (SLH-DSA) \cite{NIST_FIPS_205}.

\begin{figure}[!htbp] 
\centering
\includegraphics[width=\linewidth]{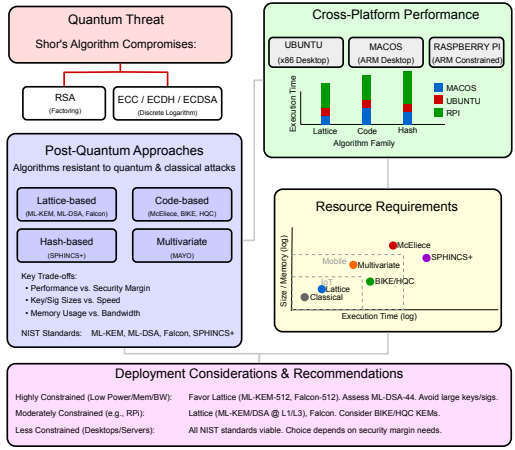}
\caption{Conceptual overview of the post-quantum cryptography landscape for consumer electronics. The emergence of quantum computers threatens classical RSA/ECC cryptography (left top). PQC offers solutions based on different mathematical foundations (left middle), each with trade-offs involving performance, security, and resource requirements (key/signature sizes, memory). This paper evaluates leading PQC candidates across diverse computing platforms, from high-performance desktops to resource-constrained devices like the Raspberry Pi (right top), analyzing their performance and resource demands (right bottom) to provide deployment considerations for different CE scenarios (bottom).}
\label{fig:conceptual_overview}
\end{figure}

Migrating the diverse CE ecosystem to PQC presents unique challenges. Many CE devices operate under strict resource constraints, including limited processing power (CPU cycles), memory (RAM), storage space, and battery life \cite{bui_health_2011, keranen_cryptographic_2014}. PQC algorithms generally exhibit different performance characteristics compared to their classical counterparts, often involving larger keys, signatures, or ciphertexts, and potentially higher computational demands \cite{paquin_benchmarking_2020}. Understanding these trade-offs is critical for selecting and deploying PQC solutions effectively in CE environments without compromising usability, functionality, or battery longevity.
Fig.~\ref{fig:conceptual_overview} provides a conceptual overview of this landscape, illustrating the quantum threat, the major PQC approaches, the varying computational environments considered, and the inherent resource trade-offs involved.

While several studies have benchmarked PQC algorithms \cite{paquin_benchmarking_2020, sikeridis_post-quantum_2020, basu_nist_2019, raavi_performance_2021}, many focus on server/desktop environments or specific algorithm classes. Fewer studies provide a comprehensive comparison of both standardized Key Encapsulation Mechanisms (KEMs) and digital signature schemes across platforms representative of the CE spectrum, from high-performance laptops/smartphones to resource-constrained embedded systems. Specifically, the impact on low-power devices typical in wearables or smart home sensors needs thorough investigation \cite{septien-hernandez_comparative_2022, kumar_securing_2022}.

This paper addresses this gap by presenting a comprehensive performance analysis of leading PQC KEMs and digital signature schemes, evaluated across multiple platforms relevant to the CE landscape: macOS (representing modern consumer laptops/desktops), Ubuntu (desktop/developer workstations), and Raspberry Pi (proxying for resource-constrained CE devices like smart home hubs, gateways, or higher-end IoT components). We compare NIST-selected/evaluated PQC algorithms against classical RSA and ECC, focusing on metrics crucial for CE:
\begin{itemize}
    \item \textbf{Execution Time:} Latency of key generation, encapsulation/decapsulation (KEMs), signing/verification (signatures), impacting user experience and energy consumption.
    \item \textbf{Memory Usage:} RAM footprint during operations, critical for low-memory devices.
    \item \textbf{Communication Costs:} Sizes of public keys, ciphertexts, and signatures, affecting bandwidth usage (e.g., OTA updates, cloud communication) and storage requirements.
\end{itemize}
We analyze performance across different NIST security levels and investigate the impact of message size on signature schemes.

The main contributions of this paper, tailored for the CE context, are:
\begin{enumerate}
    \item A cross-platform performance benchmark of standardized and candidate PQC algorithms, emphasizing implications for resource-constrained CE devices.
    \item Analysis of computational, memory, and communication overheads critical for CE design decisions (e.g., hardware selection, battery life estimation, network protocol design).
    \item Evaluation of signature performance scaling with message size, relevant for diverse CE applications from small sensor readings to large firmware updates.
    \item Specific PQC algorithm recommendations for different CE deployment scenarios (e.g., low-power wearables, smart home hubs, high-performance mobile devices) based on empirical data.
\end{enumerate}

Our findings provide CE manufacturers, software developers, and system designers with practical data and insights needed to navigate the transition to quantum-resistant security, ensuring the long-term trustworthiness and security of consumer electronic products.

The remainder of this paper is organized as follows. Section~\ref{sec:related_work} reviews existing literature on PQC benchmarking, particularly focusing on studies relevant to consumer electronics and resource-constrained environments. Section~\ref{sec:background} provides essential background on the quantum threat and the mathematical foundations of the PQC algorithm families under evaluation. Section~\ref{sec:methodology} details our experimental setup, including the hardware platforms, benchmarking framework, evaluated algorithms, and performance metrics. Section~\ref{sec:results_analysis} presents and analyzes the comprehensive performance results, comparing algorithms across platforms, security levels, and message sizes, with a focus on execution time and communication overhead, and culminates in Section~\ref{sec:deployment_recs} which synthesizes these findings into specific deployment recommendations for different classes of CE devices. Finally, Section~\ref{sec:conclusion} concludes the paper by summarizing our key findings and their implications, while Section~\ref{sec:future_work} outlines directions for future research in this critical area.

\section{Related Work}
\label{sec:related_work}
The need for post-quantum cryptography has spurred extensive research, including numerous performance evaluations. However, studies specifically addressing the broad range of CE constraints and use cases are less common.

Early benchmarking efforts, such as the SUPERCOP project \cite{bernstein_analyzing_2023}, provided foundational performance data but often focused on high-performance computing. More recent work directly evaluates NIST PQC candidates. Paquin et al. \cite{paquin_benchmarking_2020} used the `liboqs` library for benchmarking on x86 platforms, providing valuable desktop performance data but limited insights for ARM-based or heavily constrained CE devices. Sikeridis et al. \cite{sikeridis_post-quantum_2020} focused on PQC integration into TLS 1.3, analyzing handshake latency, relevant for connected CE devices, but primarily from a server/desktop perspective.

Hardware-focused evaluations, like that by Basu et al. \cite{basu_nist_2019}, examined FPGA implementations, assessing area, performance, and power. While relevant for custom silicon in CE, it doesn't fully capture software performance on general-purpose CPUs common in many CE devices.

Studies targeting resource-constrained environments are more directly related to our work. Septien-Hernandez et al. \cite{septien-hernandez_comparative_2022} provided a comparative study for IoT applications, evaluating lattice and code-based KEMs on platforms like ESP32 and Raspberry Pi. Their work highlights the feasibility challenges but didn't cover the full suite of NIST-selected algorithms or digital signatures comprehensively. Kumar et al. \cite{kumar_securing_2022} discussed securing IoT devices with PQC, emphasizing lightweight solutions. Satrya et al. \cite{satrya_comparative_2023} evaluated PQC implementations for energy systems monitoring, focusing on energy consumption on devices like Raspberry Pi, relevant for battery-powered CE, but with a limited set of algorithms. Prantl et al. \cite{prantl_performance_2021} benchmarked PQC KEMs on Cortex-M4 microcontrollers, offering insights into very low-end devices, but didn't cover signatures or higher-end CE proxies like the Raspberry Pi 4 used here.

Regarding digital signatures, Lakhan \cite{lakhan_comparative_2023} compared PQC signature algorithms concerning network performance and energy, but lacked a broad platform comparison including both KEMs and signatures. Raavi et al. \cite{raavi_performance_2021} analyzed PQC signature performance for varying message sizes on desktop platforms, which we extend to resource-constrained settings. Studies on specific families, like lattice-based schemes for embedded systems \cite{xin_vpqc_2020, fritzmann_risq-v_2020}, provide depth but not breadth across different PQC approaches.

Our work distinguishes itself by:
\begin{itemize}
    \item Providing a holistic evaluation of both NIST-standardized/round-4 KEMs and digital signature schemes.
    \item Conducting benchmarks across a spectrum of platforms representing different tiers of CE devices, from powerful (macOS, Ubuntu) to resource-constrained (Raspberry Pi 4).
    \item Analyzing a comprehensive set of metrics including execution time, memory usage, and communication costs, all critical for CE.
    \item Investigating performance scaling with security level and message size across these platforms.
    \item Deriving explicit deployment recommendations tailored to different CE scenarios based on empirical results.
\end{itemize}
This comprehensive approach aims to provide actionable guidance for integrating PQC into the diverse and often resource-limited world of consumer electronics.

\section{Background}
\label{sec:background}
This section provides an overview of quantum computing threats to classical cryptography and introduces the mathematical foundations of major post-quantum cryptographic approaches relevant to CE security.

\subsection{Quantum Computing Threats}
Classical public-key cryptography, securing countless CE devices and services, primarily relies on two hard mathematical problems: integer factorization (underpinning RSA) and the discrete logarithm problem (DLP) (underpinning Diffie-Hellman key exchange and ECC/ECDSA signatures) \cite{stallings_cryptography_2017}. Shor's algorithm, discovered in 1994 \cite{shor_polynomial-time_1999}, provides a polynomial-time quantum algorithm for both problems.

For RSA, Shor's algorithm factors a large integer $N = pq$ in $O((\log N)^3)$ time, a dramatic speedup over the best known classical algorithms which are sub-exponential \cite{pomerance_tale_1996}. The core quantum step involves finding the period $r$ of the function $f(x) = a^x \bmod N$ for a random $a < N$. If $r$ is even and $a^{r/2} \not\equiv -1 \pmod{N}$, then $\gcd(a^{r/2} \pm 1, N)$ yields factors of $N$. This effectively breaks RSA encryption and signatures. Similarly, for ECC, Shor's algorithm solves the elliptic curve discrete logarithm problem (ECDLP) $P = kG$ (find $k$ given points $P, G$) in polynomial time \cite{nielsen_quantum_2010}, compromising ECDH key exchange and ECDSA signatures widely used in mobile devices and secure web connections (TLS).

Symmetric cryptography, like AES (Advanced Encryption Standard) used for data encryption at rest and in transit within CE ecosystems, is considered more resilient. Grover's algorithm offers a quadratic speedup for searching an unstructured space of size $N$ in $O(\sqrt{N})$ time \cite{grover_fast_1996}. Applied to key search, it reduces the effective security of an $n$-bit key against quantum attack to roughly $n/2$ bits. This threat can be mitigated by doubling the key size (e.g., migrating from AES-128 to AES-256) \cite{mina_information_2021}. Consequently, the primary focus of PQC is on replacing public-key algorithms.

\subsection{Mathematical Foundations of Post-Quantum Cryptography}
PQC algorithms derive their security from mathematical problems believed to be hard for both classical and quantum computers. Major families include:

\subsubsection{Lattice-Based Cryptography}
Relies on the presumed hardness of problems defined on lattices. A lattice $\mathcal{L}$ is a discrete additive subgroup of $\mathbb{R}^n$, typically represented as the set of all integer linear combinations of a set of basis vectors $\mathbf{B} = \{\mathbf{b}_1, \dots, \mathbf{b}_k\}$:
\[ \mathcal{L}(\mathbf{B}) = \left\{ \sum_{i=1}^k z_i \mathbf{b}_i \mid z_i \in \mathbb{Z} \right\} \]
Two fundamental hard problems are:
\begin{itemize}
    \item \textbf{Shortest Vector Problem (SVP):} Given a lattice $\mathcal{L}$, find the shortest non-zero vector $\mathbf{v} \in \mathcal{L}$. Finding an exact or even approximate solution is believed hard.
    \item \textbf{Learning With Errors (LWE):} Introduced by Regev \cite{regev_lattices_2009}. Given a matrix $\mathbf{A} \in \mathbb{Z}_q^{m \times n}$, a secret vector $\mathbf{s} \in \mathbb{Z}_q^n$, and a vector $\mathbf{b} = \mathbf{A}\mathbf{s} + \mathbf{e} \pmod{q}$, where $\mathbf{e}$ is a "small" error vector sampled from a specific distribution $\chi$, distinguish samples $(\mathbf{A}, \mathbf{b})$ from uniformly random pairs $(\mathbf{A}, \mathbf{u})$. Recovering $\mathbf{s}$ is the computational variant. LWE is proven to be at least as hard as worst-case lattice problems like GapSVP \cite{regev_lattices_2009}.
\end{itemize}
Structured variants like Ring-LWE (RLWE) and Module-LWE (MLWE) operate over polynomial rings (e.g., $R_q = \mathbb{Z}_q[X]/\langle \Phi_n(X) \rangle$) or modules over these rings, offering better efficiency and smaller key sizes \cite{lyubashevsky_toolkit_2013}.
NIST selected ML-KEM (based on Kyber \cite{bos_crystals-kyber_2018}, using MLWE) and ML-DSA (based on Dilithium \cite{ducas_crystals-dilithium_2018}, using MLWE) as primary standards, with ML-DSA now specified in FIPS 204 \cite{NIST_FIPS_204}. Falcon \cite{fouque_falcon_2018}, another standard, uses NTRU lattices \cite{bernstein_ntru_2016}.
For instance, a simplified MLWE-based KEM (like Kyber) involves:
\begin{itemize}
    \item \textbf{KeyGen:} Sample matrix $\mathbf{A} \in R_q^{k \times k}$, small vectors $\mathbf{s}, \mathbf{e} \in R_q^k$. Public key $\mathsf{pk} = (\mathbf{A}, \mathbf{t} = \mathbf{A}\mathbf{s} + \mathbf{e})$. Private key $\mathsf{sk} = \mathbf{s}$.
    \item \textbf{Encaps:} Sample small $\mathbf{r}, \mathbf{e}_1 \in R_q^k, e_2 \in R_q$. Compute $\mathbf{u} = \mathbf{A}^\top \mathbf{r} + \mathbf{e}_1$ and $v = \mathbf{t}^\top \mathbf{r} + e_2 + \lfloor \frac{q}{2} \rfloor \cdot m$ (where $m$ is the message bit). Ciphertext $\mathsf{ct} = (\mathbf{u}, v)$. Shared secret derived from $\mathbf{r}$ and $\mathsf{pk}$.
    \item \textbf{Decaps:} Compute $m' = v - \mathbf{s}^\top \mathbf{u}$. If $m'$ is close to $\lfloor \frac{q}{2} \rfloor$, output $m=1$, else $m=0$. Shared secret derived from $\mathbf{s}$ and $\mathsf{ct}$.
\end{itemize}
The security relies on the MLWE assumption. These schemes are attractive for CE due to their relatively good performance and moderate key/ciphertext sizes.

\subsubsection{Code-Based Cryptography}
Based on the hardness of decoding general linear error-correcting codes. Given a generator matrix $\mathbf{G} \in \mathbb{F}_q^{k \times n}$ or a parity-check matrix $\mathbf{H} \in \mathbb{F}_q^{(n-k) \times n}$ of a linear code $C$, and a received word $\mathbf{y} = \mathbf{c} + \mathbf{e}$ where $\mathbf{c} \in C$ and $\mathbf{e}$ is an error vector of weight $t$, the core problem is finding $\mathbf{c}$ (or $\mathbf{e}$).
\begin{itemize}
    \item \textbf{Syndrome Decoding Problem:} Given $\mathbf{H}$, a syndrome $\mathbf{s} = \mathbf{H}\mathbf{y}^\top$, and weight $t$, find $\mathbf{e}$ of weight $t$ such that $\mathbf{H}\mathbf{e}^\top = \mathbf{s}$. This is NP-hard for random linear codes \cite{berlekamp_inherent_2003}.
\end{itemize}
The McEliece cryptosystem \cite{mceliece_public-key_1978} uses a class of codes (e.g., binary Goppa codes) with an efficient decoding algorithm $D$.
\begin{itemize}
    \item \textbf{KeyGen:} Choose a Goppa code with $(n, k, t)$ parameters and generator matrix $\mathbf{G}$. Choose a random non-singular matrix $\mathbf{S} \in \mathbb{F}_2^{k \times k}$ and permutation matrix $\mathbf{P} \in \mathbb{F}_2^{n \times n}$. Public key $\mathsf{pk} = \mathbf{G'} = \mathbf{S}\mathbf{G}\mathbf{P}$. Private key $\mathsf{sk} = (\mathbf{S}, \mathbf{G}, \mathbf{P})$ (or equivalent information allowing efficient decoding $D$).
    \item \textbf{Encrypt:} To encrypt message $\mathbf{m} \in \mathbb{F}_2^k$, choose random error vector $\mathbf{e} \in \mathbb{F}_2^n$ of weight $t$. Ciphertext $\mathbf{c} = \mathbf{m}\mathbf{G'} + \mathbf{e}$.
    \item \textbf{Decrypt:} Compute $\mathbf{c'} = \mathbf{c}\mathbf{P}^{-1} = \mathbf{m}\mathbf{S}\mathbf{G} + \mathbf{e}\mathbf{P}^{-1}$. Use decoder $D$ to correct errors $\mathbf{e'} = \mathbf{e}\mathbf{P}^{-1}$ yielding $\mathbf{m}\mathbf{S}\mathbf{G}$. Compute $\mathbf{m}\mathbf{S} = (\mathbf{m}\mathbf{S}\mathbf{G})\mathbf{G}^{-1}_{\text{right}}$. Recover $\mathbf{m} = (\mathbf{m}\mathbf{S})\mathbf{S}^{-1}$.
\end{itemize}
Classic McEliece \cite{baldi_post-quantum_2017} is a NIST finalist known for strong security confidence but suffers from very large public keys ($\mathbf{G'}$ is $k \times n$, where $k, n$ are thousands), posing challenges for CE devices. Other schemes like BIKE \cite{aragon_bike_2022} (using QC-MDPC codes) and HQC \cite{melchor_hamming_2018} aim for smaller keys.

\subsubsection{Hash-Based Signatures}
Security relies solely on the cryptographic properties of a hash function $H: \{0,1\}^* \to \{0,1\}^n$, primarily collision resistance and preimage resistance. Lamport's one-time signature (OTS) scheme \cite{lamport_constructing_1979} is foundational.
\begin{itemize}
    \item \textbf{KeyGen (Lamport):} For an $m$-bit message, generate $2m$ random secret values $\{x_{i,0}, x_{i,1}\}_{i=1..m}$. Compute public key components $y_{i,j} = H(x_{i,j})$. $\mathsf{sk} = \{x_{i,j}\}$, $\mathsf{pk} = \{y_{i,j}\}$.
    \item \textbf{Sign (Lamport):} For message $M=(M_1...M_m)$, the signature is $\sigma = (x_{1, M_1}, x_{2, M_2}, \dots, x_{m, M_m})$.
    \item \textbf{Verify (Lamport):} Check if $H(\sigma_i) = y_{i, M_i}$ for all $i=1..m$.
\end{itemize}
This is one-time secure. Schemes like Winternitz OTS (WOTS+) \cite{buchmann_security_2011} improve signature size. Stateful schemes like XMSS \cite{hulsing_xmss_2018} use Merkle trees to combine many OTS keys under one public key, but require careful state management. Stateless schemes like SPHINCS+ \cite{bernstein_sphincs_2019} use hypertrees and few-time signatures (FORS \cite{kudinov_sphincs_2022}) to avoid state, simplifying deployment, crucial for robust CE implementations. SPHINCS+ was chosen by NIST as a standard, now specified as SLH-DSA in FIPS 205 \cite{NIST_FIPS_205}. Hash-based signatures offer strong security assurance but often have large signature sizes (tens of KB) and can be slower for signing compared to lattice-based schemes.

\subsubsection{Multivariate Cryptography}
Based on the hardness of solving systems of $m$ multivariate polynomial equations in $n$ variables over a finite field $\mathbb{F}_q$: Find $\mathbf{x} = (x_1, \dots, x_n)$ such that $p_i(x_1, \dots, x_n) = y_i$ for $i=1..m$, given polynomials $p_i$ and target values $y_i$. This MQ problem is NP-hard \cite{garey_computers_2002}.
Signature schemes typically use a trapdoor structure: an easily invertible central map $\mathcal{F}: \mathbb{F}_q^n \to \mathbb{F}_q^m$ (often quadratic) composed with secret affine maps $\mathcal{S}: \mathbb{F}_q^n \to \mathbb{F}_q^n$ and $\mathcal{T}: \mathbb{F}_q^m \to \mathbb{F}_q^m$.
\begin{itemize}
    \item \textbf{KeyGen:} Choose $\mathcal{S}, \mathcal{F}, \mathcal{T}$. Public key $\mathsf{pk} = \mathcal{P} = \mathcal{T} \circ \mathcal{F} \circ \mathcal{S}$. Private key $\mathsf{sk} = (\mathcal{S}, \mathcal{F}, \mathcal{T})$.
    \item \textbf{Sign:} To sign a document hash $h$, compute target $\mathbf{y} = \mathcal{T}^{-1}(h)$. Compute $\mathbf{z} = \mathcal{F}^{-1}(\mathbf{y})$ using the central map's trapdoor. Signature $\sigma = \mathcal{S}^{-1}(\mathbf{z})$.
    \item \textbf{Verify:} Check if $\mathcal{P}(\sigma) = h$.
\end{itemize}
Schemes like Rainbow \cite{ding_multivariate_2006} (broken \cite{beullens_breaking_2022}) and Multivariate Algebraic Signatures Overfield
(MAYO) \cite{beullens_mayo_2022} (based on Oil and Vinegar) differ in their choice of $\mathcal{F}$ and transformations. They can offer very short signatures, potentially attractive for specific CE use cases, but history shows vulnerability to algebraic attacks.

\subsubsection{Isogeny-Based Cryptography}
Uses the presumed hardness of problems related to isogenies between elliptic curves over finite fields. An isogeny $\phi: E \to E'$ is a non-constant morphism between elliptic curves that is also a group homomorphism. The core problem involves finding an isogeny $\phi$ given $E$ and $E' = \phi(E)$, potentially with additional information about $\phi$ (like its degree or action on torsion points).
SIKE (Supersingular Isogeny Key Encapsulation) \cite{jao_sike_2017} used walks in the supersingular isogeny graph.
\begin{itemize}
    \item \textbf{Setup:} Public curve $E_0/\mathbb{F}_{p^2}$.
    \item \textbf{KeyGen (Alice):} Choose secret isogeny $\phi_A: E_0 \to E_A$. Public key $\mathsf{pk}_A = E_A$.
    \item \textbf{KeyGen (Bob):} Choose secret isogeny $\phi_B: E_0 \to E_B$. Public key $\mathsf{pk}_B = E_B$.
    \item \textbf{Shared Secret (Alice):} Compute $E_{BA} = \phi_A(E_B)$. J-invariant $j(E_{BA})$ determines secret.
    \item \textbf{Shared Secret (Bob):} Compute $E_{AB} = \phi_B(E_A)$. J-invariant $j(E_{AB})$ determines secret. ($j(E_{AB}) = j(E_{BA})$).
\end{itemize}
SIKE offered very small key sizes but was broken by attacks exploiting auxiliary torsion point information \cite{castryck_efficient_2023}. Research into other potentially secure isogeny constructions continues.

\subsection{Comparison of Post-Quantum Approaches}
Table~\ref{tab:pqc_comparison} summarizes key characteristics relevant to CE deployment.

\begin{table*}[!ht]
\centering
\caption{Comparison of Post-Quantum Cryptographic Approaches}
\label{tab:pqc_comparison}
\footnotesize
\begin{tabular*}{\linewidth}{@{\extracolsep{\fill}}p{2.2cm}p{3cm}p{3cm}p{3cm}p{3cm}@{}}
\toprule
\textbf{Property} & \textbf{Lattice-based} & \textbf{Code-based} & \textbf{Hash-based (Stateless)} & \textbf{Multivariate} \\
\midrule
Mathematical Foundation & LWE, SVP over lattices & Syndrome Decoding Problem & Collision-resistant hash functions & Solving multivariate equations \\
\midrule
Hard Problem Studied Since & 1996 (lattices), 2005 (LWE) \cite{regev_lattices_2009} & 1978 (McEliece) \cite{singh_code_2020, mceliece_public-key_1978} & 1979 (Lamport sigs.) \cite{lamport_constructing_1979} & 1988 (Matsumoto-Imai) \cite{ding_new_2004}\\
\midrule
Key Size (Relative) & Small-Medium & Large (McEliece), Medium (BIKE/HQC) & Small (public), Large (private internal state representation) & Small-Medium \\
\midrule
Sig./Ciphertext Size (Relative) & Small-Medium & Small (McEliece), Medium (BIKE/HQC) & Large & Small-Medium (Signatures) \\
\midrule
Comp. Efficiency & Generally High & Medium-High (Encryption/Decryption), Slow KeyGen (McEliece) & Medium/Slow (Signing), Fast (Verification) & Fast (Verification), Medium/Slow (Signing) \\
\midrule
Quantum Security Confidence & Medium-High (Active research area) & High (Long history) & Very High (Relies on hash functions) & Medium (Subject to algebraic attacks) \\
\midrule
CE Relevance & Strong candidates due to balance & Challenging (McEliece keys), Potential (BIKE/HQC) & Good for firmware sigs (verify often), bandwidth concern & Potential for short signatures if security holds \\
\midrule
NIST Standards/ Candidates & ML-KEM, ML-DSA, Falcon & Classic McEliece (Round 4), BIKE, HQC & SPHINCS+ (Standard) & MAYO (Round 4) \\
\midrule
Side-Channel Risk & Potential in implementations \cite{aydin_horizontal_2021} & Considered lower risk & Considered lower risk & Potential in implementations \\
\bottomrule
\end{tabular*}
\end{table*}

Lattice-based cryptography (ML-KEM, ML-DSA, Falcon) emerges as a strong contender for general-purpose CE use due to its balance of performance and size \cite{ducas_crystals-dilithium_2018, bos_crystals-kyber_2018}. Code-based schemes like Classic McEliece face hurdles due to large keys \cite{septien-hernandez_comparative_2022}. Hash-based signatures (SPHINCS+) offer robust security but large signatures impact bandwidth and potentially storage on constrained devices \cite{bernstein_sphincs_2019}. Multivariate schemes remain interesting for potentially small signatures but require ongoing security scrutiny. Our experiments quantify these trade-offs across relevant platforms.


\section{Methodology}
\label{sec:methodology}
\subsection{Experimental Setup}
Our performance evaluation was conducted across three distinct hardware platforms chosen to represent a spectrum of capabilities found in the consumer electronics ecosystem:

\begin{itemize}
    \item \textbf{macOS (High-End Consumer/Development):} An Apple Mac mini equipped with an M4 chip (10-core ARM-based CPU, 10-core GPU), 16GB unified RAM, running macOS Sequoia (15.4). This platform represents modern high-performance consumer devices like laptops, powerful tablets, or developer workstations.
    \item \textbf{Ubuntu (Desktop/Development):} A desktop system with an 11th Gen Intel Core i7-11700 CPU (8 cores/16 threads @ 2.50GHz base, x86-64 architecture), 32GB DDR4 RAM, running Ubuntu 22.04 LTS. This serves as a common desktop/server environment and provides an x86-64 performance comparison point.
    \item \textbf{Raspberry Pi 4 Model B (Resource-Constrained CE Proxy):} Featuring a Broadcom BCM2711 SoC with a quad-core ARM Cortex-A72 (ARMv8) 64-bit CPU clocked at 1.5GHz, 4GB LPDDR4 RAM, running Raspberry Pi OS (64-bit Debian Bullseye based). This platform acts as a proxy for moderately resource-constrained CE devices such as smart home hubs, advanced IoT gateways, automotive infotainment units, or set-top boxes, offering insights into performance on common embedded ARM architectures under tighter resource budgets.
\end{itemize}
This platform selection allows us to analyze PQC performance not just in ideal desktop conditions but also under limitations more typical of deployed CE hardware.

\subsection{Benchmarking Framework and Libraries}
The benchmarking framework was implemented using Python 3.\footnote{The source code for the benchmarking framework and analysis scripts used in this study is publicly available at: \url{https://github.com/dcommey/pqc_evaluation}}
\begin{itemize}
    \item \textbf{PQC Algorithms:} We utilized the `liboqs` open-source C library (version 0.8.0 via the `oqs-python` 0.8.0 wrapper) \cite{stebila_post-quantum_2017}, which provides implementations of numerous PQC schemes submitted to the NIST standardization process. `liboqs` was compiled on each platform using standard toolchains (GCC 11.4 on Ubuntu/Raspberry Pi, Apple Clang 14.0.3 on macOS) with the `-O3` optimization level enabled.
    \item \textbf{Classical Algorithms:} Baseline classical cryptography performance was measured using the Python `cryptography` library (version 41.0.7), which relies on the underlying OpenSSL library (version 3.3.1) available on each platform for optimized implementations of RSA, ECDH, ECDSA, and Ed25519.
\end{itemize}
Execution times for cryptographic operations were measured using high-resolution system timers available through Python. To ensure statistical stability and minimize transient effects, each operation was typically repeated 1000 times within the benchmarking loop (fewer iterations were used for exceptionally long operations, such as Classic McEliece key generation). The initial runs were discarded as warm-up, and the mean and standard deviation were calculated from the subsequent, stable measurements.

\subsection{Data Consolidation and Naming Standardization}
Raw performance results were collected independently from each platform. For clarity and consistency with the evolving cryptographic landscape and official NIST standards, a data consolidation and naming standardization process was applied. This process addressed instances where algorithms were potentially benchmarked under different names but corresponded to the same underlying NIST standard specification. Specifically:

\begin{itemize}
    \item Both historical \texttt{Kyber} submission variants and official \texttt{ML-KEM} standard variants were present in the raw results. The analysis script identified the standard \texttt{ML-KEM} designation for each corresponding parameter set (e.g., identifying both \texttt{Kyber512} and \texttt{ML-KEM-512} as representing the same standard, ML-KEM-512). To prevent duplication, where results existed under both names for the same platform and parameter set, the entry originating from the historical \texttt{Kyber} name was excluded from the final dataset.
    \item Similarly, both historical \texttt{Dilithium} submission variants and official \texttt{ML-DSA} standard variants were present. The script identified the standard \texttt{ML-DSA} designation corresponding to each historical \texttt{Dilithium} level (e.g., mapping \texttt{Dilithium2} to ML-DSA-44, \texttt{Dilithium3} to ML-DSA-65, \texttt{Dilithium5} to ML-DSA-87). Where results existed under both the historical \texttt{Dilithium} name (after mapping to the corresponding ML-DSA level) and the official \texttt{ML-DSA} name for the same platform and effective parameter set, the entry originating from the historical \texttt{Dilithium} name was excluded.
\end{itemize}

Through this consolidation process, the official NIST standard names (\texttt{ML-KEM}, \texttt{ML-DSA}) were preferentially retained, ensuring that only one unique entry per standard algorithm variant per platform was used in the subsequent analysis and discussion. This approach allows for direct comparison based on the final standardized algorithms. These standardized names were then used to assign algorithms to families (e.g., Lattice, Code, Hash) based on their underlying mathematical structure.

\subsection{Evaluated Algorithms}
We evaluated a comprehensive set of KEMs and digital signature schemes, encompassing NIST standards, prominent Round 4 candidates, and classical baselines for comparison. The specific algorithms are listed below, using the consolidated naming where applicable.

\textbf{Key Encapsulation Mechanisms (KEMs):}
\begin{itemize}
    \item \textit{Lattice-based (MLWE):} ML-KEM (512, 768, 1024) [NIST Standard]
    \item \textit{Lattice-based (LWE):} FrodoKEM (640-AES, 640-SHAKE, 976-AES, 976-SHAKE, 1344-AES, 1344-SHAKE)
    \item \textit{Lattice-based (NTRU-like):} sntrup761 (NTRU Prime variant)
    \item \textit{Code-based:} BIKE (L1, L3, L5), Classic McEliece (348864f, 460896f, 6688128f, 6960119f, 8192128f) [f-variants used], HQC (128, 192, 256)
    \item \textit{Classical:} RSA (2048, 3072, 4096 bits), ECDH (P-256, P-384, P-521)
\end{itemize}

\textbf{Digital Signature Schemes:}
\begin{itemize}
 \item \textit{Lattice-based (MLWE):} ML-DSA (44, 65, 87) [NIST Standard \cite{NIST_FIPS_204}]
 \item \textit{Lattice-based (NTRU-like):} Falcon (512, 1024) [NIST Standard]
 \item \textit{Hash-based:} SPHINCS+ (variants combining SHA2/SHAKE, 128/192/256 security levels, and f/s trade-offs) [NIST Standard \cite{NIST_FIPS_205}]
 \item \textit{Multivariate:} MAYO (1, 2, 3, 5)
 \item \textit{Classical:} RSA (2048, 3072, 4096 bits), ECDSA (P-256, P-384, P-521), Ed25519
\end{itemize}
Parameter sets targeting NIST security levels 1, 3, and 5 were included where available.

\subsection{Performance Metrics}
The evaluation focused on metrics most relevant to CE deployment:

\begin{enumerate}
    \item \textbf{Execution Time:} Wall-clock time (in milliseconds, ms) for core cryptographic operations: Key Generation, Encapsulation, Decapsulation for KEMs; Key Generation, Signing, Verification for signatures. We report mean and standard deviation across iterations.
    \item \textbf{Communication Costs:} Sizes (in bytes) of cryptographic artifacts that need to be stored or transmitted: Public Keys, Ciphertexts (for KEMs), and Signatures (for signature schemes). Shared secret sizes (KEMs) and secret key sizes were also recorded for completeness.
    \item \textbf{Message Size Impact (Signatures):} Execution times for signing and verification were measured using varying message sizes: \SI{1}{\kilo\byte}, \SI{10}{\kilo\byte}, \SI{100}{\kilo\byte}, and \SI{1}{\mega\byte} (\SI{1048576}{\byte}). This assesses how performance scales with data size for applications like firmware updates or data logging.
    \item \textbf{Resource Requirement Trade-offs:} We analyze the relationship between computational cost (approximated by total operation time) and communication/storage cost (approximated by combined key/ciphertext or key/signature size) to visualize the resource trade-off space, particularly on the constrained Raspberry Pi platform.
\end{enumerate}
While peak memory usage (RAM footprint) is crucial for CE, reliable and cross-platform measurement within the software framework proved inconsistent across platforms and libraries. Therefore, this analysis primarily focuses on execution time and communication/storage size, using the latter as an indirect indicator of memory pressure associated with cryptographic object handling.

\section{Results and Analysis}
\label{sec:results_analysis}
This section presents the performance benchmark results derived from the methodology described above, analyzing the implications for deploying PQC in consumer electronics. We focus on the consolidated data, prioritizing NIST standard names (ML-KEM, ML-DSA). The macOS platform serves as the \texttt{REFERENCE\_PLATFORM} for general performance plots, while the Raspberry Pi serves as the \texttt{RESOURCE\_PLATFORM} for constrained analysis, unless otherwise specified.

\subsection{Performance Comparison Across Platforms}
Comparing performance across macOS, Ubuntu, and Raspberry Pi reveals the significant impact of hardware capabilities on PQC execution.

\begin{figure}[!ht]
\centering
\includegraphics[width=0.9\linewidth]{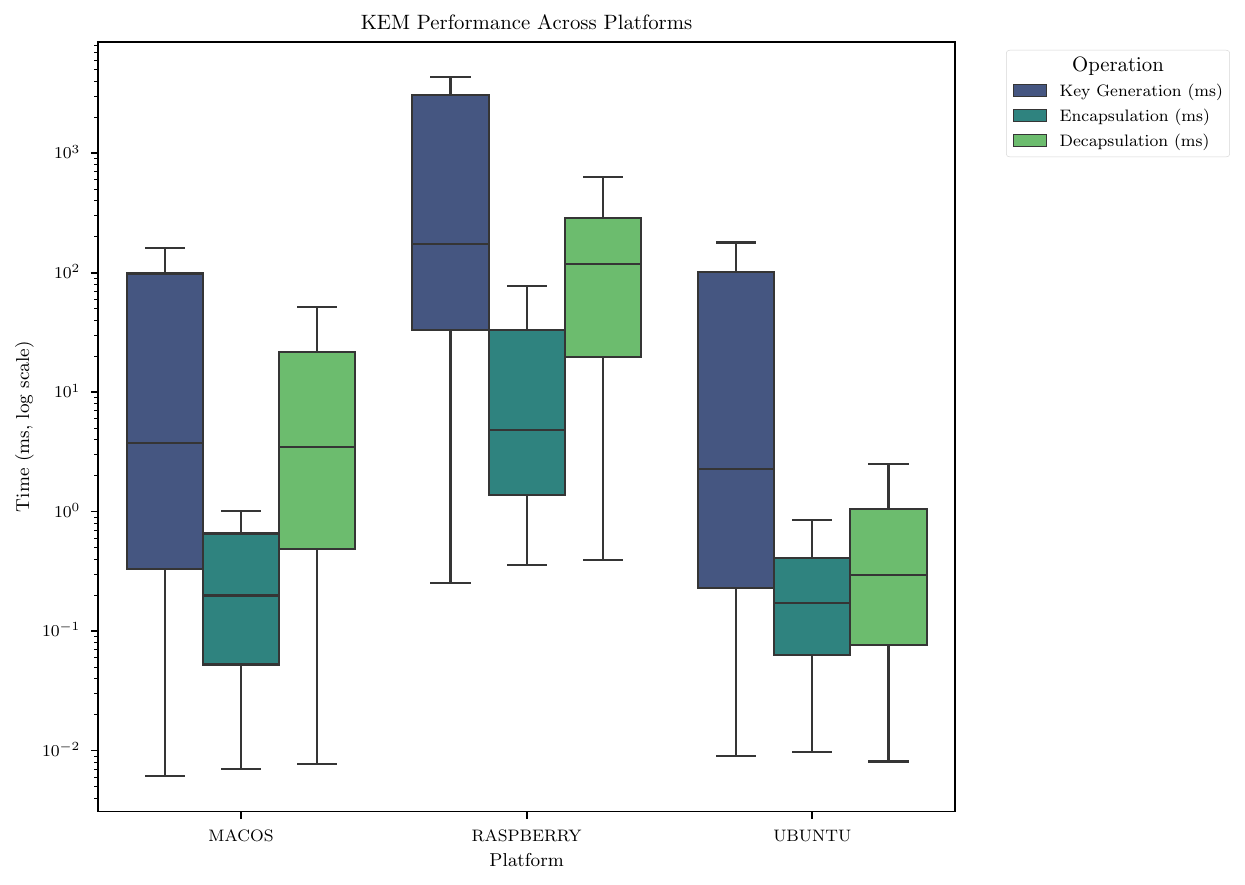} 
\caption{KEM Performance Across Platforms (Execution Time, ms - Log Scale). Compares key generation, encapsulation, and decapsulation times. Note the logarithmic scale accentuating the significant performance gap on the Raspberry Pi.}
\label{fig:kem_platform_perf}
\end{figure}

Figure~\ref{fig:kem_platform_perf} illustrates the execution times for KEM operations. As expected, the Raspberry Pi is considerably slower than the desktop platforms (macOS, Ubuntu). macOS (M4 ARM) and Ubuntu (Intel x86) exhibit closer performance, though variations exist depending on the algorithm family and specific operation.

Table~\ref{tab:kem_platform_stats} quantifies these differences, averaging across algorithm types. On average, PQC KEM operations are roughly 45-50 times slower on the Raspberry Pi compared to macOS for encapsulation/decapsulation, while key generation slowdown varies more widely depending on the algorithm mix (including very slow McEliece vs. fast lattices). Classical KEMs also slow down, but PQC operations, particularly decapsulation, show a larger relative penalty on the constrained platform.

\begin{table*}[!ht]
\centering
\caption{Average KEM Performance Statistics Across Platforms (ms)}
\label{tab:kem_platform_stats}
\footnotesize
\begin{tabular*}{\linewidth}{@{\extracolsep{\fill}}lrrrrrrr@{}}
\toprule
Platform & Type & Key MEAN & Key STD & Encapsulation MEAN & Encapsulation STD & Decapsulation MEAN & Decapsulation STD \\
\midrule
MACOS & Classical & 102.18 & 156.54 & 0.22 & 0.20 & 2.25 & 2.69 \\
MACOS & Post-Quantum & 54.52 & 87.35 & 0.71 & 1.03 & 15.04 & 18.07 \\
RASPBERRY & Classical & 2533.90 & 4113.17 & 2.69 & 2.34 & 34.15 & 45.74 \\
RASPBERRY & Post-Quantum & 2017.67 & 3679.47 & 36.28 & 57.31 & 245.85 & 237.86 \\
UBUNTU & Classical & 91.26 & 149.63 & 0.16 & 0.11 & 0.74 & 0.78 \\
UBUNTU & Post-Quantum & 40.09 & 57.59 & 1.37 & 3.30 & 1.99 & 4.96 \\
\bottomrule
\end{tabular*}
\end{table*}

\begin{figure}[!ht]
\centering
\includegraphics[width=0.9\linewidth]{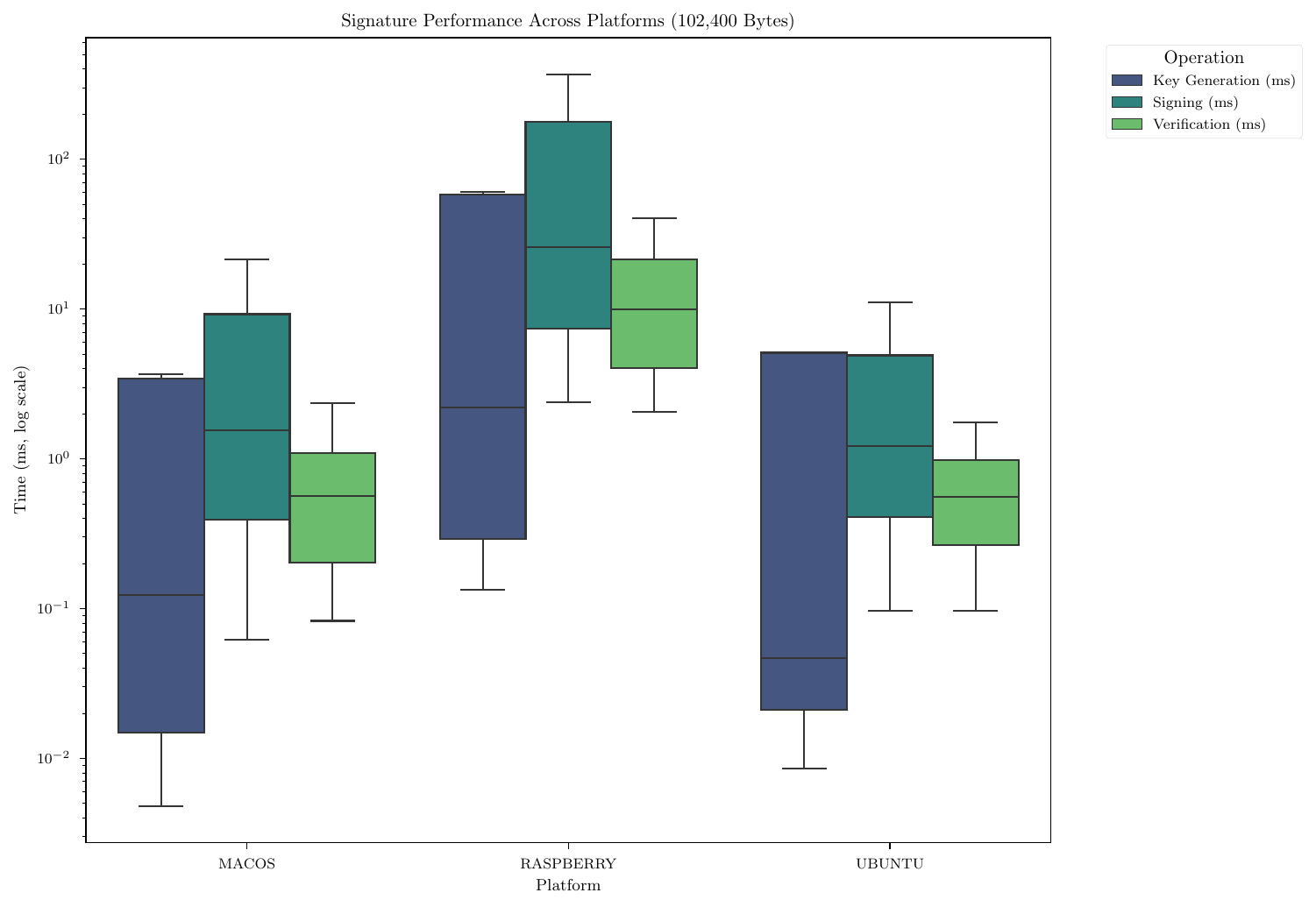} 
\caption{Signature Performance Across Platforms (\SI{100}{\kilo\byte} Message, ms - Log Scale). Demonstrates the performance hierarchy across platforms for signature operations.}
\label{fig:sig_platform_perf_102400}
\end{figure}

Signature schemes show a similar trend. Figure~\ref{fig:sig_platform_perf_102400} displays performance for a \SI{100}{\kilo\byte} message size. Verification times generally see the largest relative increase on the Raspberry Pi compared to signing or key generation for many PQC schemes.

Detailed tables summarizing signature statistics for various message sizes (\SI{1}{\kilo\byte}, \SI{10}{\kilo\byte}, \SI{100}{\kilo\byte}, \SI{1}{\mega\byte}) are provided in the Appendix. Table~\ref{tab:sig_platform_stats_102400} shows the summary for the \SI{100}{\kilo\byte} message case. These confirm substantial slowdowns for signatures on the Raspberry Pi, often exceeding 20x compared to macOS, particularly for computationally intensive operations or specific algorithms like SPHINCS+ signing. This performance differential is a primary consideration for selecting algorithms for battery-powered or real-time sensitive CE applications.

\begin{table*}[!ht]
\centering
\caption{Average Signature Performance Statistics Across Platforms (\SI{100}{\kilo\byte} Message, ms)}
\label{tab:sig_platform_stats_102400}
\footnotesize
\begin{tabular*}{\linewidth}{@{\extracolsep{\fill}}lrrrrrrr@{}}
\toprule
Platform & Type & Key MEAN & Key STD & Signing MEAN & Signing STD & Verification MEAN & Verification STD \\
\midrule
MACOS & Classical & 89.98 & 154.00 & 1.98 & 2.67 & 0.29 & 0.25 \\
MACOS & Post-Quantum & 9.35 & 21.25 & 86.85 & 204.78 & 0.80 & 0.63 \\
RASPBERRY & Classical & 2286.47 & 4173.19 & 30.84 & 43.99 & 4.66 & 3.29 \\
RASPBERRY & Post-Quantum & 180.00 & 403.26 & 1668.83 & 3851.65 & 15.10 & 11.00 \\
UBUNTU & Classical & 78.60 & 141.72 & 0.72 & 0.78 & 0.23 & 0.16 \\
UBUNTU & Post-Quantum & 5.34 & 11.22 & 44.23 & 105.62 & 0.83 & 0.66 \\
\bottomrule
\end{tabular*}
\end{table*}

\subsection{Desktop Platform Comparison (macOS vs. Ubuntu)}
Analyzing the performance ratios between the two desktop platforms (Ubuntu/macOS) reveals nuances beyond raw speed. Detailed ratio plots and summary tables are provided in the Appendix.

Generally, performance is comparable for many lattice-based schemes (ML-KEM, ML-DSA), with ratios often close to 1.0. However, architecture-specific optimizations or library differences can lead to variations. For instance, our Ubuntu/x86 platform showed an advantage for FrodoKEM and NTRU-Prime, while macOS/ARM favoured ML-KEM and code-based schemes. Signature ratios also showed some variability. This implies that developers targeting cross-platform CE applications (e.g., companion apps on desktops, different OS-based CE devices) should consider potential performance variations even between relatively powerful platforms.

\subsection{Performance on Resource-Constrained Devices (Raspberry Pi)}
The performance impact on the Raspberry Pi is crucial for understanding PQC feasibility in constrained CE. Slowdown factors relative to macOS quantify this penalty (see Appendix Figures~\ref{fig:kem_ratio_raspberry_app} and \ref{fig:sig_ratio_raspberry_app}, and Appendix Table~\ref{tab:appendix_kem_platform_ratios}).

As highlighted in the Appendix KEM ratio summary table (Table~\ref{tab:appendix_kem_platform_ratios}), KEMs exhibit significant slowdowns. Lattice schemes (ML-KEM, NTRU-Prime) slow down by roughly 30-50x. Code-based schemes vary, typically ~40x for key generation/encapsulation but less for decapsulation. FrodoKEM variants show substantial slowdowns, often exceeding 100x.

Signatures face similar challenges (see Appendix Signature ratio summary table, Table~\ref{tab:sig_platform_ratios_app}). Lattice signatures (ML-DSA, Falcon) typically slow down by 15-25x. Hash-based SPHINCS+ signing is drastically slower (often >100x), while verification slowdown is less severe (~25x). Multivariate and Randomized Subset Difference Problem (RSDP) schemes also see significant slowdowns. Classical algorithms generally exhibit smaller slowdown factors (10-25x) but are cryptographically insecure long-term.

These slowdowns directly translate to increased latency and energy consumption, potentially impacting CE usability and battery life significantly.

\subsection{Resource Requirement Trade-offs}
\begin{figure}[!ht]
\centering
\includegraphics[width=0.9\linewidth]{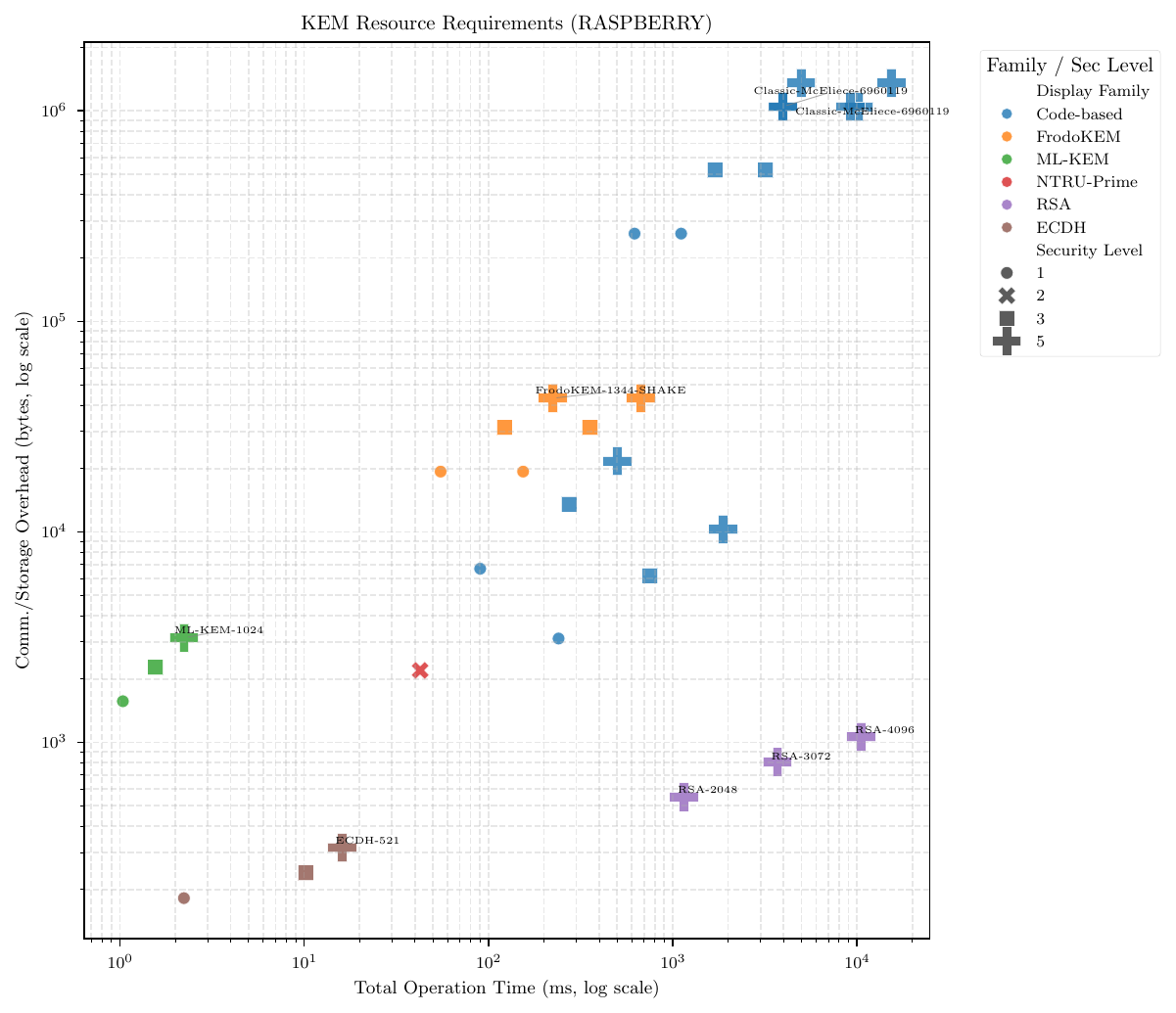} 
\caption{KEM Resource Requirements (Time vs. Comm./Storage Overhead) on Raspberry Pi (Log Scales). Ideal algorithms are in the bottom-left. ML-KEM variants offer excellent balance. Classic McEliece requires excessive communication/storage size. ECDH is most compact and relatively fast.}
\label{fig:kem_resource_requirements_raspberry}
\end{figure}

Optimizing for CE often involves balancing computational cost (time) against memory/storage/communication cost (size). Figure~\ref{fig:kem_resource_requirements_raspberry} plots total KEM operation time (KeyGen+Encaps+Decaps) against communication size (PubKey+Ciphertext) on the Raspberry Pi.

ML-KEM variants stand out, residing in the desirable bottom-left quadrant (low time, low size). Classical ECDH is also very efficient. NTRU Prime (sntrup761) is similar in size to ML-KEM but slower. Code-based schemes and FrodoKEM occupy the middle ground in both time and size. Classic McEliece demonstrates the extreme trade-off: very large size requirements but moderate operational time (dominated by slow key generation, not shown in this combined operational metric).

\begin{figure}[!ht]
\centering
\includegraphics[width=0.9\linewidth]{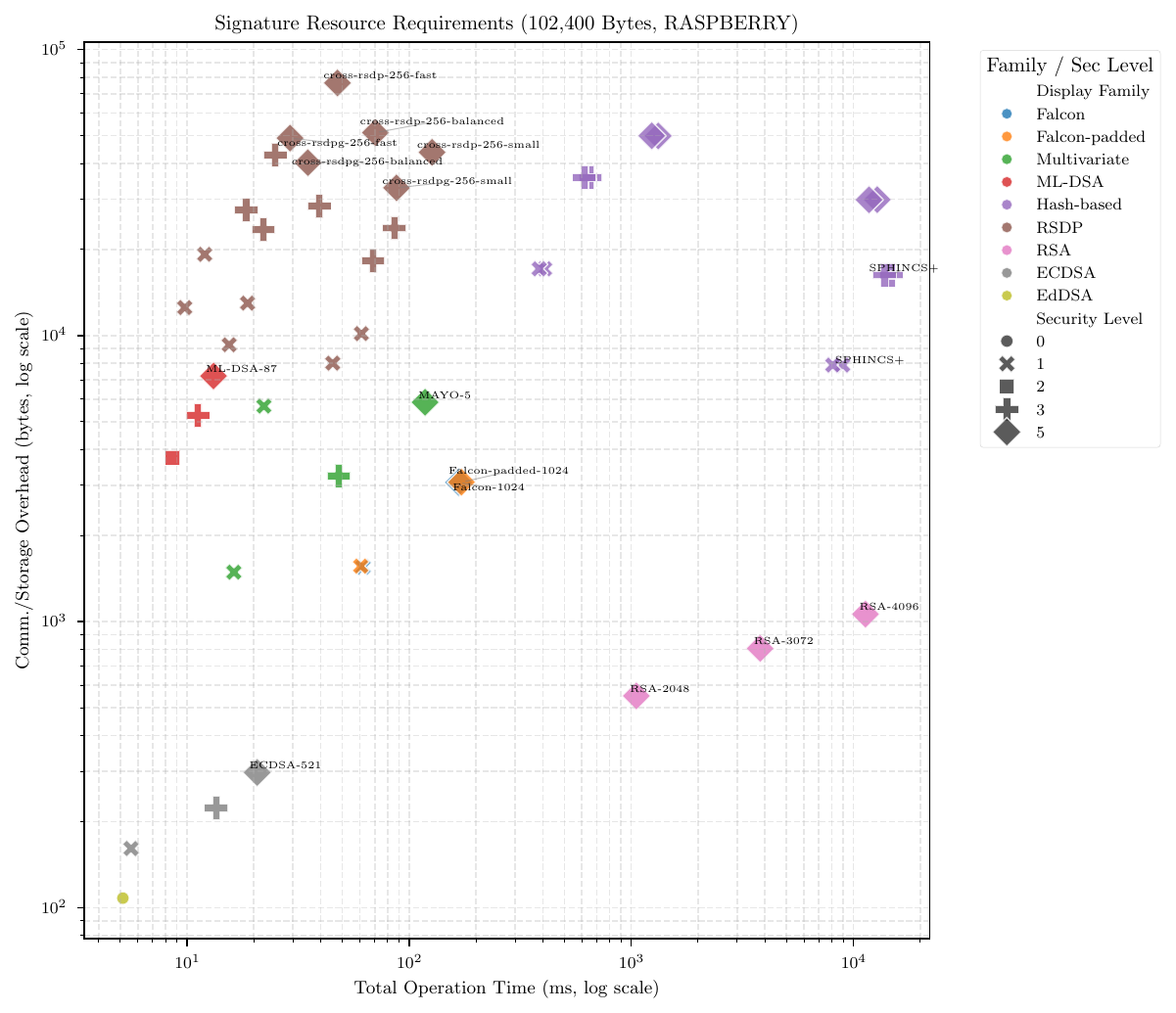} 
\caption{Signature Resource Requirements (Time vs. Comm./Storage Overhead) on Raspberry Pi (\SI{100}{\kilo\byte} Message, Log Scales). Falcon offers small sizes and moderate time. ML-DSA is balanced. SPHINCS+ variants require large sizes and have very slow total operational times (dominated by signing). Classical ECC/EdDSA are highly efficient but insecure.}
\label{fig:signature_resource_requirements_raspberry}
\end{figure}

Figure~\ref{fig:signature_resource_requirements_raspberry} presents the trade-off for signatures (Total Time vs. PubKey+Signature Size) on the Pi, using the \SI{100}{\kilo\byte} message results.

Falcon excels in compactness (small keys and signatures) with moderate total time. ML-DSA offers a good balance slightly larger than Falcon. Multivariate schemes (like MAYO) have small signatures but require larger keys and significant time. SPHINCS+ variants have very small public keys but enormous signatures and slow total times (signing dominates). RSDP signatures are also very large. Classical EdDSA and ECDSA remain the most resource-efficient but are vulnerable. These plots visually guide algorithm selection based on which resource (time, size/memory/bandwidth) is the primary constraint for a given CE device.

\subsection{Communication Overhead}
\begin{figure}[!ht]
\centering
\includegraphics[width=0.9\linewidth]{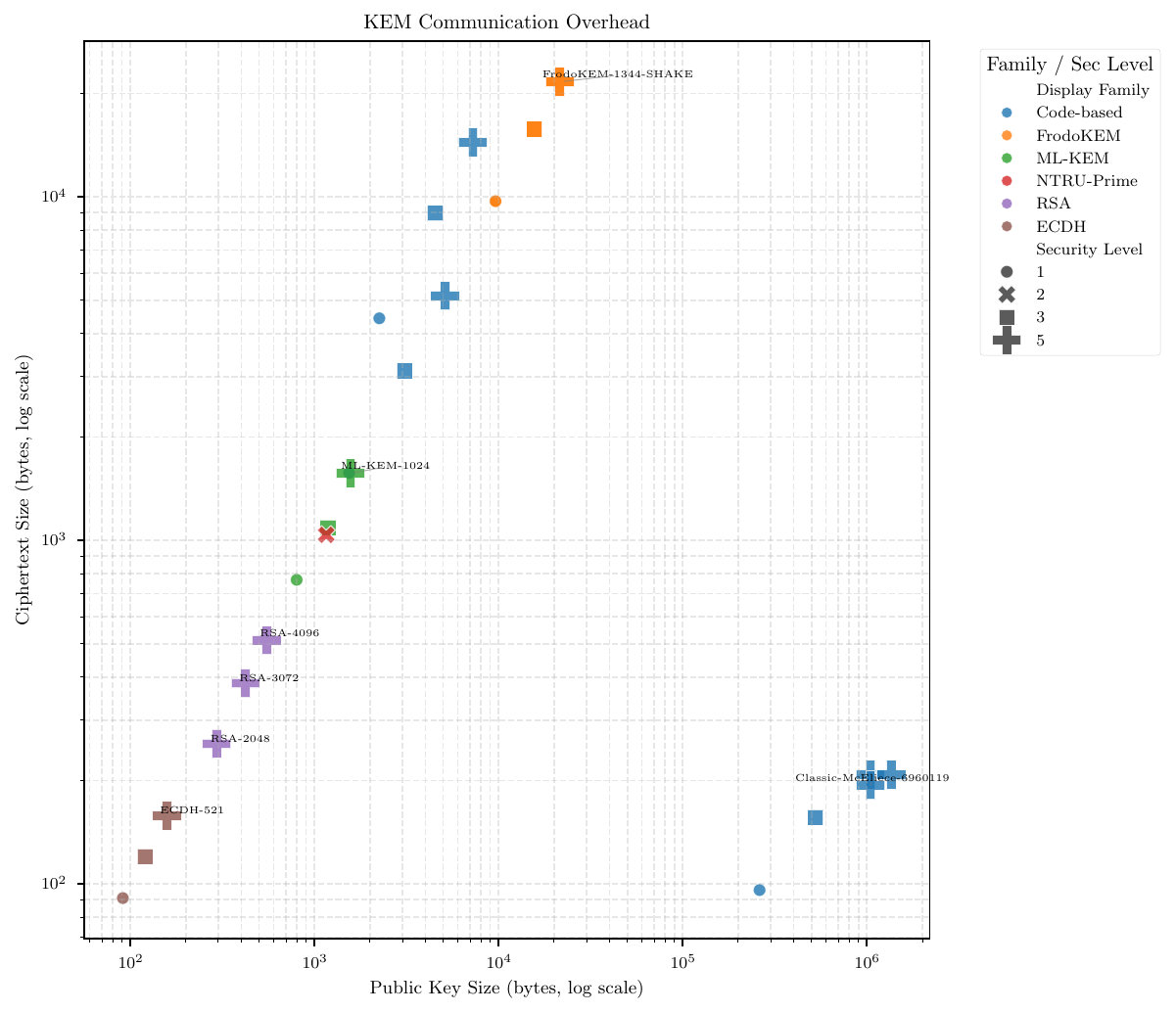}
\caption{KEM Communication Overhead (Public Key vs. Ciphertext Size, Bytes - Log Scale, MACOS data). Highlights the vast range from compact ECDH/ML-KEM to enormous Classic McEliece public keys.}
\label{fig:kem_communication_overhead}
\end{figure}

Focusing specifically on the size of transmitted data, Figure~\ref{fig:kem_communication_overhead} plots public key size versus ciphertext size for KEMs (using macOS data for clarity of labels).

Table~\ref{tab:kem_comm_costs} provides average sizes per family based on the algorithms tested. Code-based schemes (dominated by Classic McEliece in our tested set) show huge public keys but small ciphertexts. ML-KEM offers sizes around \SI{1}{\kilo\byte} to \SI{1.5}{\kilo\byte} for both, suitable for many network protocols. FrodoKEM requires significantly larger sizes (tens of kilobytes combined).

\begin{table}[!ht]
\centering
\footnotesize
\setlength{\tabcolsep}{2pt}
\caption{Average KEM communication costs by family (bytes)}
\label{tab:kem_comm_costs}
\resizebox{\linewidth}{!}{%
\begin{tabular}{@{}ll*{3}{r}*{3}{r}*{3}{r}@{}}
\toprule
\multicolumn{2}{@{}l}{\textbf{Family / Type}} & \multicolumn{3}{c}{\textbf{Public‑key}} & \multicolumn{3}{c}{\textbf{Cipher‑text}} & \multicolumn{3}{c@{}}{\textbf{Shared secret}} \\
\cmidrule(lr){3-5}\cmidrule(lr){6-8}\cmidrule(lr){9-11}
& & Mean & Min & Max & Mean & Min & Max & Mean & Min & Max \\ \midrule
Code‑based & PQ & 387\,198 & 1\,541 & 1\,357\,824 & 3\,503 & 96 & 14\,421 & 41 & 32 & 64 \\
ECDH & Classical & 123 & 91 & 158 & 123 & 91 & 158 & 49 & 32 & 66 \\
FrodoKEM & PQ & 15\,589 & 9\,616 & 21\,520 & 15\,699 & 9\,720 & 21\,632 & 24 & 16 & 32 \\
ML‑KEM & PQ & 1\,184 & 800 & 1\,568 & 1\,141 & 768 & 1\,568 & 32 & 32 & 32 \\
NTRU‑Prime & PQ & 1\,158 & 1\,158 & 1\,158 & 1\,039 & 1\,039 & 1\,039 & 32 & 32 & 32 \\
RSA & Classical & 422 & 294 & 550 & 384 & 256 & 512 & 32 & 32 & 32 \\ \bottomrule
\end{tabular}}%
\end{table}

\begin{figure}[!ht]
\centering
\includegraphics[width=0.9\linewidth]{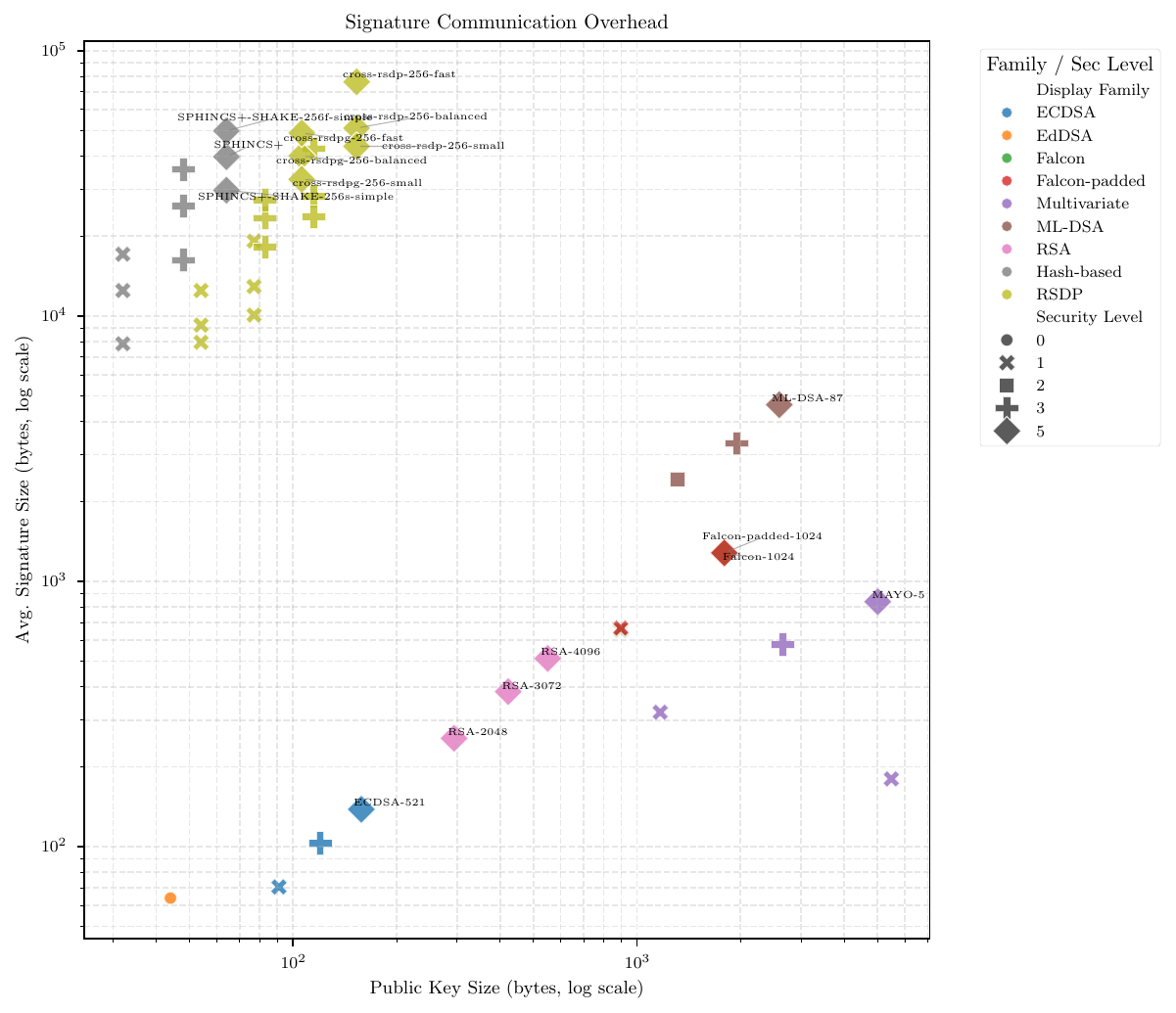} 
\caption{Signature Communication Overhead (Public Key vs. Avg. Signature Size, Bytes - Log Scale, MACOS data). Hash-based/RSDP variants have large signatures but small keys. Falcon is compact in both dimensions. ML-DSA is moderate.}
\label{fig:signature_communication_overhead}
\end{figure}

For signatures (Figure~\ref{fig:signature_communication_overhead} and Table~\ref{tab:sig_comm_costs}), Hash-based (SPHINCS+) and RSDP clearly dominate in signature size (tens of kilobytes), while Falcon (\SI{<1}{\kilo\byte} to \SI{1.3}{\kilo\byte}) and Multivariate (MAYO) (\SI{<1}{\kilo\byte}) are notably compact. ML-DSA (\SI{2.4}{\kilo\byte} to \SI{4.6}{\kilo\byte}) represents a middle ground. Public key sizes are generally smaller for signatures compared to KEMs, with SPHINCS+ and RSDP having very small public keys.

\begin{table}[!ht]
\centering
\footnotesize
\setlength{\tabcolsep}{2pt}
\caption{Average Signature Communication Costs by Family (bytes)}
\label{tab:sig_comm_costs}
\resizebox{\columnwidth}{!}{%
\begin{tabular}{@{}ll*{3}{r}*{3}{r}@{}}
\toprule
\multicolumn{2}{@{}l}{\textbf{Family / Type}} & \multicolumn{3}{c}{\textbf{Public Key Size}} & \multicolumn{3}{c@{}}{\textbf{Signature Size}} \\
\cmidrule(lr){3-5} \cmidrule(lr){6-8}
& & Mean & Min & Max & Mean & Min & Max \\
\midrule
ECDSA & Classical & 123 & 91 & 158 & 104 & 70 & 138 \\
EdDSA & Classical & 44 & 44 & 44 & 64 & 64 & 64 \\
Falcon & Post-Quantum & 1\,345 & 897 & 1\,793 & 964 & 656 & 1\,273 \\
Falcon-padded & Post-Quantum & 1\,345 & 897 & 1\,793 & 973 & 666 & 1\,280 \\
Hash-based & Post-Quantum & 48 & 32 & 64 & 26\,080 & 7\,856 & 49\,856 \\
ML-DSA & Post-Quantum & 1\,952 & 1\,312 & 2\,592 & 3\,452 & 2\,420 & 4\,627 \\
Multivariate & Post-Quantum & 3\,580 & 1\,168 & 5\,488 & 479 & 180 & 838 \\
RSA & Classical & 422 & 294 & 550 & 384 & 256 & 512 \\
RSDP & Post-Quantum & 98 & 54 & 153 & 29\,338 & 7\,956 & 76\,298 \\
\bottomrule
\end{tabular}}
\end{table}

These sizes directly impact CE applications. Large Hash-based or RSDP signatures might strain low-bandwidth networks (BLE, LoRaWAN) or consume excessive space in secure boot partitions. Code-based KEM keys might require complex key distribution mechanisms for constrained devices. Lattice schemes offer a generally well-regarded balance for typical CE network interactions and storage capabilities.

\subsection{Performance Across Different Security Levels}
\begin{figure}[!ht]
\centering
\includegraphics[width=0.9\linewidth]{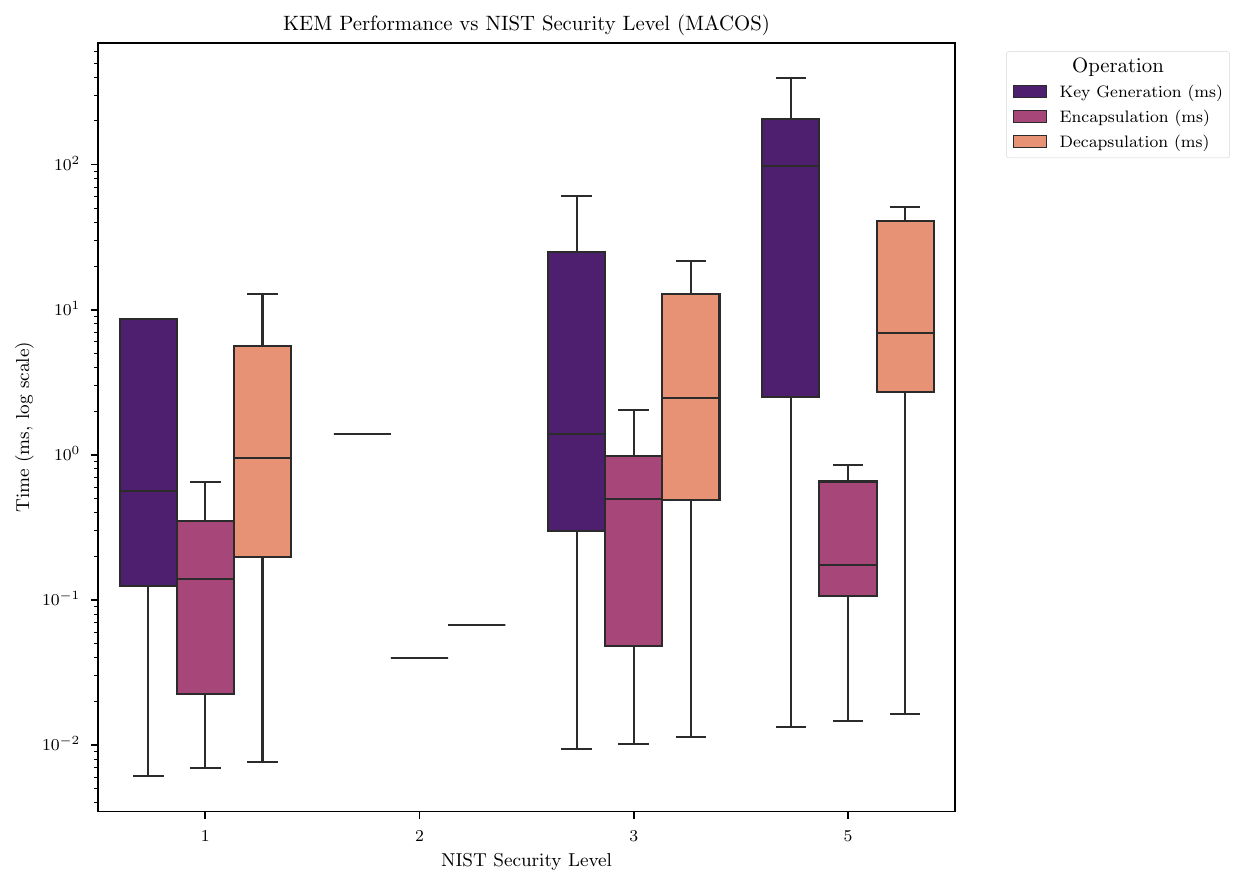}
\caption{KEM Performance vs. NIST Security Level (MACOS). Shows average operation times for each security level.}
\label{fig:kem_security_analysis}
\end{figure}

\begin{figure}[!ht]
\centering
\includegraphics[width=0.9\linewidth]{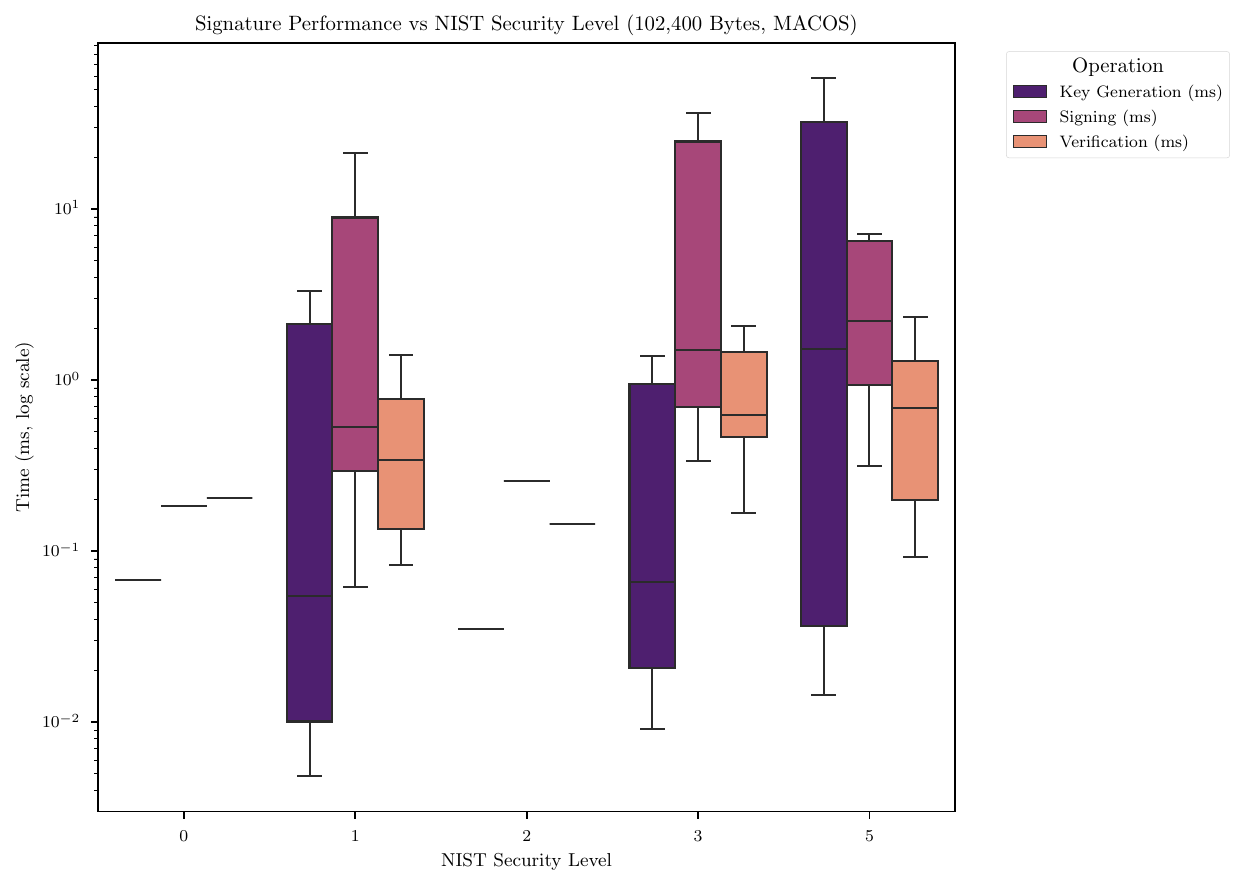}
\caption{Signature Performance vs. NIST Security Level (\SI{100}{\kilo\byte} Message, MACOS). Illustrates scaling trends for signature operations.}
\label{fig:sig_security_analysis_102400}
\end{figure}

Choosing the right security level involves trade-offs. Figures~\ref{fig:kem_security_analysis} and \ref{fig:sig_security_analysis_102400} (using \SI{100}{\kilo\byte} messages for signatures) illustrate how average PQC performance scales compared to classical options across NIST Levels 1, 3, and 5 on the reference macOS platform. Table~\ref{tab:security_impact} provides quantitative averages combining performance and size metrics.

Generally, higher security levels incur costs: longer execution times and larger cryptographic objects.
\begin{itemize}
    \item \textbf{Time Scaling:} PQC operation times tend to increase moderately with security level, often less dramatically than RSA key generation. Lattice schemes scale relatively well.
    \item \textbf{Size Scaling:} PQC public key and ciphertext/signature sizes show more significant increases. As seen in Table~\ref{tab:security_impact}, average PQC signature sizes increase substantially from Level 1 to Level 5, driven largely by Hash-based and RSDP schemes. Average PQC KEM key/ciphertext sizes also grow considerably, influenced by code-based and FrodoKEM parameters at higher levels. Lattice schemes exhibit more moderate size growth.
\end{itemize}
For CE devices with long expected lifetimes or handling sensitive data, targeting Level 3 or 5 security is advisable. Designers must weigh the increased computational and communication costs against the required security margin.

\begin{table*}[!ht]
\centering
\caption{Impact of security level on average performance and size (macOS)}
\label{tab:security_impact}
\resizebox{\linewidth}{!}{%
\begin{tabular}{@{}l*{18}{r}@{}}
\toprule
& \multicolumn{2}{c}{Ciphertext size (bytes)}
& \multicolumn{2}{c}{Decapsulation (ms)}
& \multicolumn{2}{c}{Encapsulation (ms)}
& \multicolumn{2}{c}{Key‑generation (ms)}
& \multicolumn{4}{c}{Public‑key size (bytes)}
& \multicolumn{2}{c}{Signature size (bytes)}
& \multicolumn{2}{c}{Signing (ms)}
& \multicolumn{2}{c@{}}{Verification (ms)}\\
\cmidrule(lr){2-3}\cmidrule(lr){4-5}\cmidrule(lr){6-7}\cmidrule(lr){8-9}
\cmidrule(lr){10-13}\cmidrule(lr){14-15}\cmidrule(lr){16-17}\cmidrule(lr){18-19}
& Classical & PQ & Classical & PQ & Classical & PQ & Classical & PQ
& Classical & Classical & PQ & PQ   
& Classical & PQ
& Classical & PQ
& Classical & PQ\\
Security level & KEM & KEM & KEM & KEM & KEM & KEM & KEM & KEM
& KEM & Sig & KEM & Sig   
& Sig & Sig
& Sig & Sig
& Sig & Sig\\
\midrule
0 & – & – & – & – & – & – & – & – & – & 44.00 & – & – & 64.00 & – & 0.38 & – & 0.31 & –\\
1 & 91.00 & 3772.29 & 0.05 & 4.41 & 0.05 & 0.31 & 0.02 & 10.28 & 91.00 & 91.00 & 78008.86 & 687.62 & 70.00 & 8897.08 & 0.12 & 49.03 & 0.14 & 0.72\\
2 & – & 1039.00 & – & 0.07 & – & 0.04 & – & 1.40 & – & – & 1158.00 & 1312.00 & – & 2420.00 & – & 0.45 & – & 0.33\\
3 & 120.00 & 6425.86 & 0.49 & 8.38 & 0.50 & 0.75 & 0.20 & 28.95 & 120.00 & 120.00 & 155481.86 & 486.00 & 103.00 & 23177.82 & 0.39 & 103.65 & 0.68 & 1.09\\
5 & 327.50 & 5966.09 & 3.25 & 27.41 & 0.20 & 1.00 & 153.22 & 103.77 & 356.00 & 356.00 & 632476.82 & 935.00 & 322.50 & 33098.62 & 3.37 & 81.80 & 0.33 & 1.19\\
\bottomrule
\end{tabular}}%
\end{table*}

\subsection{Impact of Message Size on Signature Performance}
\begin{figure}[htbp]
  \centering
  \begin{subfigure}{\linewidth}
    \includegraphics[width=\linewidth,
                     trim=5pt 5pt 25pt 15pt,clip]%
                     {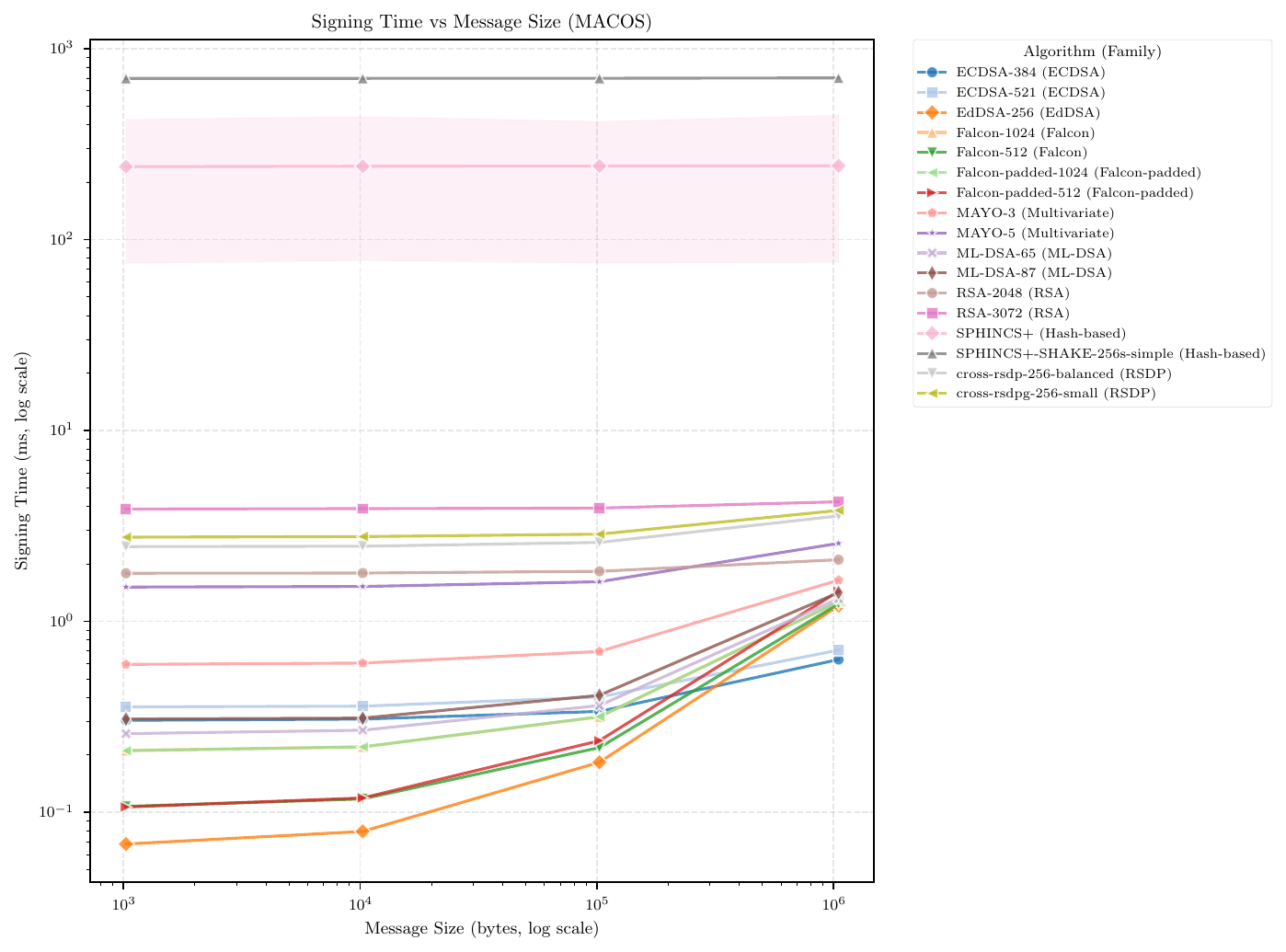}
    \caption{Signing time vs.\ message size on macOS}
    \label{fig:msg_size_impact_signing}
  \end{subfigure}

  \vspace{0.6em} 

  \begin{subfigure}{\linewidth}
    \includegraphics[width=\linewidth,
                     trim=5pt 5pt 25pt 15pt,clip]%
                     {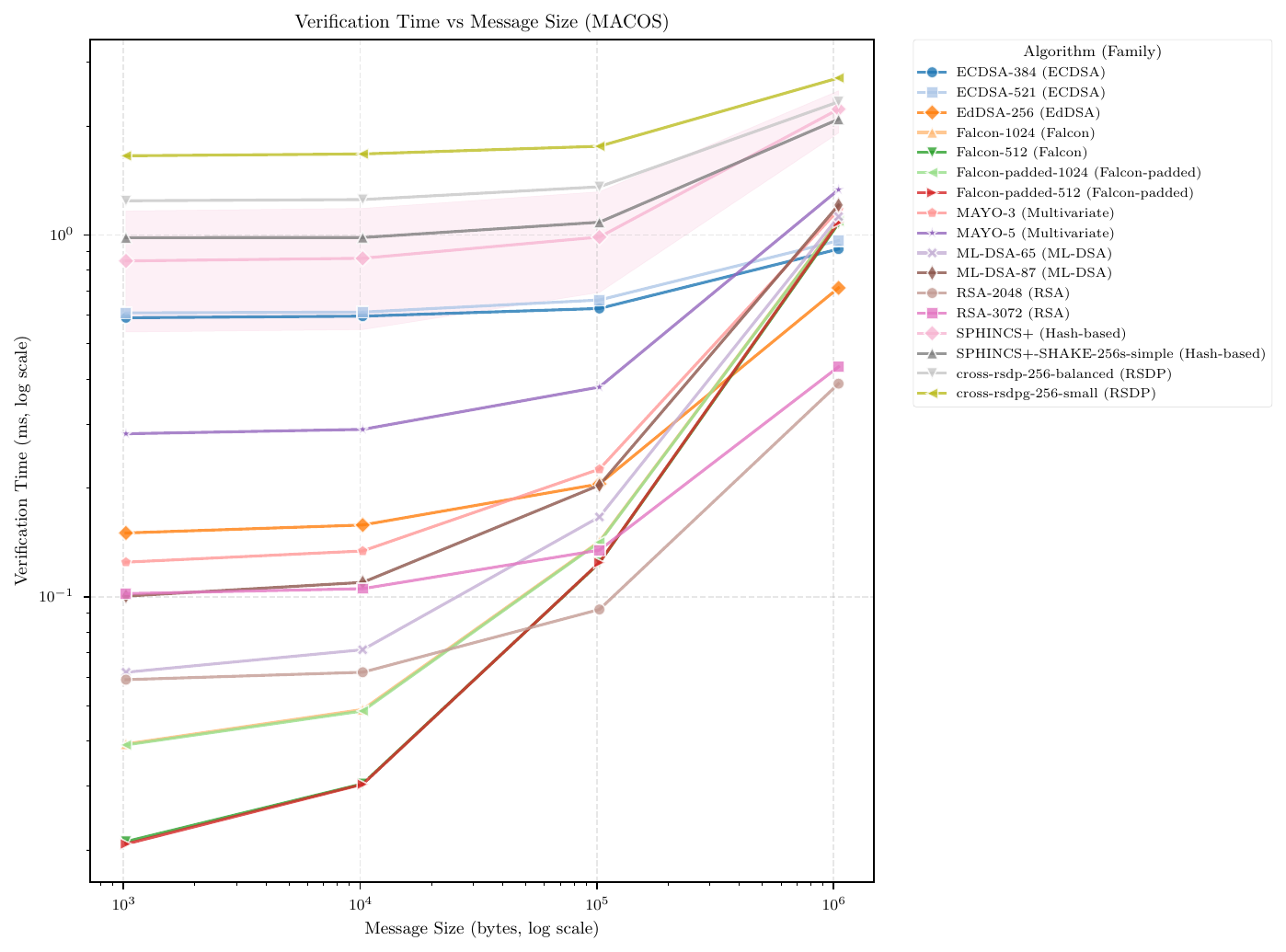}
    \caption{Verification time vs.\ message size on macOS}
    \label{fig:msg_size_impact_verification}
  \end{subfigure}

  \caption{Impact of message size on signature performance (log scales).  
           Selected algorithms across message sizes from 1 kB to 1 MB.  
           SPHINCS\textsuperscript{+} stays almost flat because it hashes
           the message only once, whereas lattice schemes, especially during
           verification, grow with~$\log$ (message size).}
  \label{fig:msg_size_impact}
\end{figure}

The performance of signing and verification can depend on the size of the message being processed. Figures~\ref{fig:msg_size_impact_signing} and \ref{fig:msg_size_impact_verification} plot operation times against message size for selected representative algorithms on macOS.

Key observations regarding scaling (relative to a 1KB baseline):
\begin{itemize}
    \item \textbf{Hash-based Stability (SPHINCS+):} Performance is nearly independent of message size (scaling factors close to 1.0), as the primary cost involves hashing the message once initially. This makes it predictable for applications involving large data blobs, like firmware updates.
    \item \textbf{Lattice Scaling:} ML-DSA and Falcon show noticeable performance scaling, particularly for verification, as message size increases. Verification times can increase significantly when moving from \SI{1}{\kilo\byte} to \SI{1}{\mega\byte} messages. Signing scales less dramatically. This is because the message digest interacts with the core lattice operations.
    \item \textbf{Classical Scaling:} RSA and ECDSA verification also scale with message size due to hashing, but the base times are often lower for smaller messages. EdDSA (Ed25519) shows noticeable scaling for signing large messages.
    \item \textbf{Other PQC:} Multivariate and RSDP schemes generally show scaling similar to lattice schemes, as they also involve hashing the message.
\end{itemize}
For CE applications frequently signing or verifying large data blocks, the scaling behavior is a critical factor. Hash-based schemes are advantageous here if their other overheads are acceptable; otherwise, the scaling limits of other PQC schemes must be considered.

\subsection{Algorithm Family Comparisons}
\begin{figure}[!htbp]
\centering
\includegraphics[width=\linewidth]{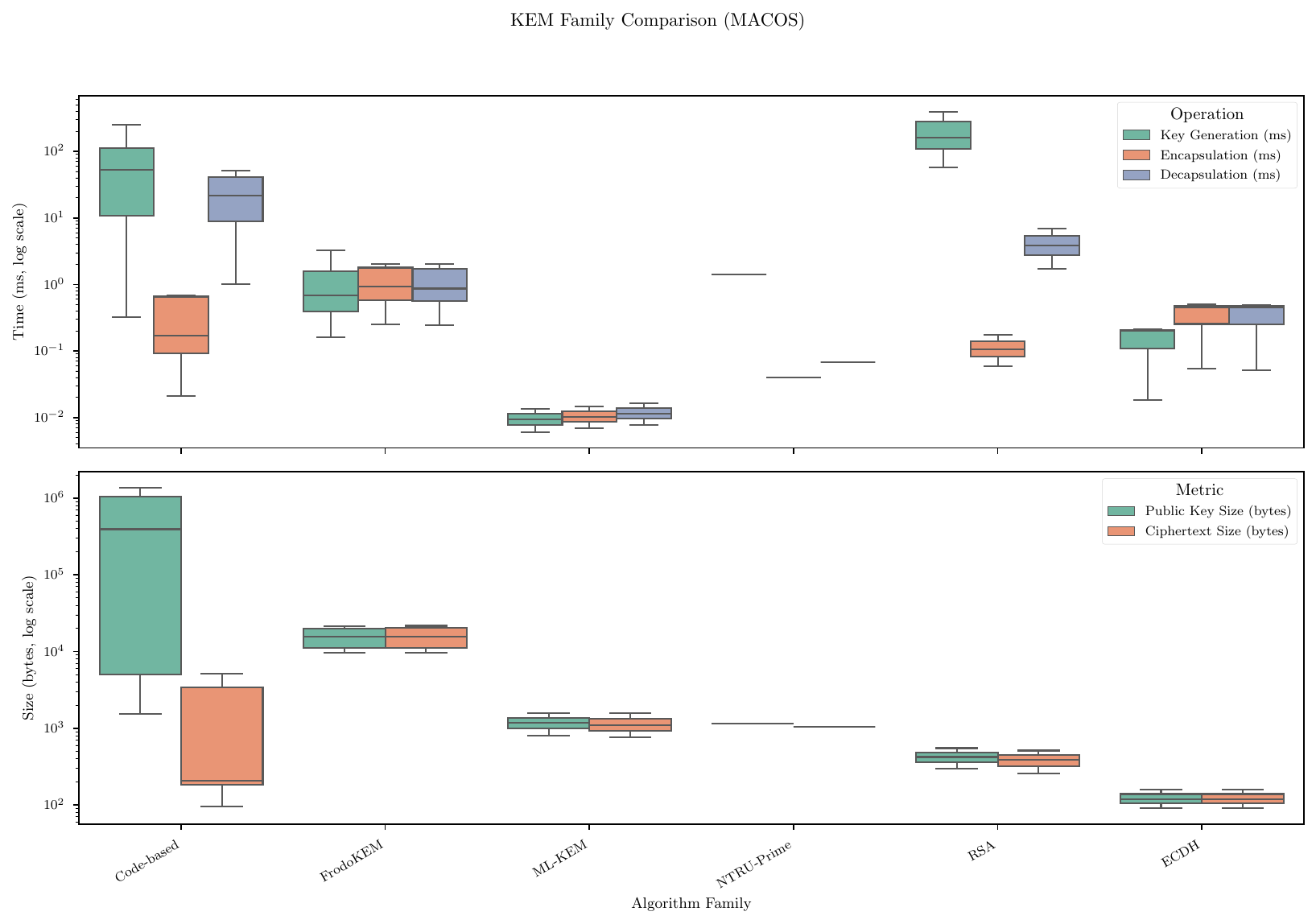}
\caption{KEM Family Comparison (MACOS). Top: Operation Times (ms, Log Scale). Bottom: Communication Sizes (bytes, Log Scale). ML-KEM shows excellent speed and moderate size. Code-based schemes have large key sizes. FrodoKEM is slower with larger sizes.}
\label{fig:kem_family_analysis}
\end{figure}

Aggregating results provides a high-level view of family characteristics. Figure~\ref{fig:kem_family_analysis} compares KEM families based on average performance and size on macOS (see also Table~\ref{tab:kem_family_summary}).

\begin{table*}[!htbp]
\centering
\caption{KEM family summary statistics (macOS)}
\label{tab:kem_family_summary}
\resizebox{\linewidth}{!}{%
\begin{tabular}{@{}l*{20}{r}@{}}
\toprule
& \multicolumn{4}{c}{Key‑generation (ms)} & \multicolumn{4}{c}{Encapsulation (ms)} & \multicolumn{4}{c}{Decapsulation (ms)} & \multicolumn{4}{c}{Public‑key size (bytes)} & \multicolumn{4}{c@{}}{Cipher‑text size (bytes)}\\
\cmidrule(lr){2-5}\cmidrule(lr){6-9}\cmidrule(lr){10-13}\cmidrule(lr){14-17}\cmidrule(lr){18-21}
Family & mean & std & min & max & mean & std & min & max & mean & std & min & max & mean & std & min & max & mean & std & min & max\\
\midrule
Code‑based   & 88.07 & 97.95 & 0.32 & 285.45 & 0.63 & 0.95 & 0.02 & 3.47 & 23.93 & 18.00 & 1.00 & 51.34 & 530\,912.00 & 531\,741.34 & 1\,541 & 1\,357\,824 & 2\,462.38 & 4\,084.65 & 96 & 14\,421\\
ECDH         & 0.14 & 0.11 & 0.02 & 0.21 & 0.34 & 0.25 & 0.05 & 0.50 & 0.33 & 0.25 & 0.05 & 0.49 & 123.00 & 33.60 & 91 & 158 & 123.00 & 33.60 & 91 & 158\\
FrodoKEM     & 1.16 & 1.18 & 0.16 & 3.25 & 1.39 & 1.28 & 0.25 & 3.68 & 1.36 & 1.28 & 0.24 & 3.66 & 15\,589.33 & 5\,323.73 & 9\,616 & 21\,520 & 15\,698.67 & 5\,327.32 & 9\,720 & 21\,632\\
ML‑KEM       & 0.01 & 0.00 & 0.01 & 0.01 & 0.01 & 0.00 & 0.01 & 0.01 & 0.01 & 0.00 & 0.01 & 0.02 & 1\,184.00 & 384.00 & 800 & 1\,568 & 1\,141.33 & 402.66 & 768 & 1\,568\\
NTRU‑Prime   & 1.40 & – & 1.40 & 1.40 & 0.04 & – & 0.04 & 0.04 & 0.07 & – & 0.07 & 0.07 & 1\,158.00 & – & 1\,158 & 1\,158 & 1\,039.00 & – & 1\,039 & 1\,039\\
RSA          & 204.22 & 173.27 & 56.84 & 395.10 & 0.11 & 0.06 & 0.06 & 0.17 & 4.17 & 2.63 & 1.73 & 6.96 & 422.00 & 128.00 & 294 & 550 & 384.00 & 128.00 & 256 & 512\\
\bottomrule
\end{tabular}}%
\end{table*}

\begin{figure}[!htbp]
\centering
\includegraphics[width=\linewidth]{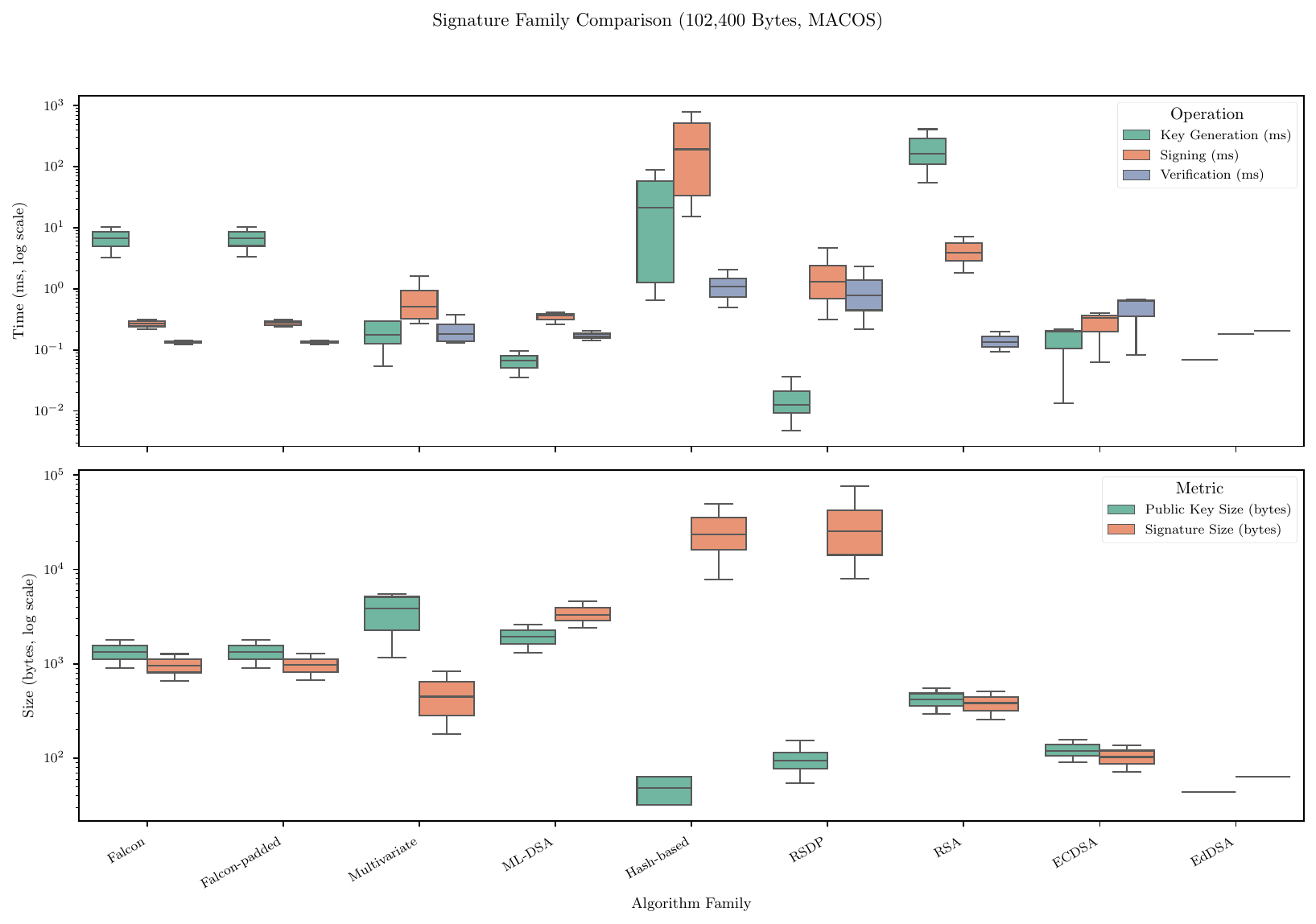} 
\caption{Signature Family Comparison (\SI{100}{\kilo\byte} Message, MACOS). Top: Operation Times (ms, Log Scale). Bottom: Communication Sizes (bytes, Log Scale). Falcon offers fast verification and small signatures. ML-DSA provides balanced performance and size. Hash-based has slow signing, fast verification, and very large signatures. Multivariate shows slow signing. RSDP has very large signatures.}
\label{fig:sig_family_analysis_102400}
\end{figure}

Figure~\ref{fig:sig_family_analysis_102400} compares signature families (using \SI{100}{\kilo\byte} messages on macOS), with statistics summarized in Table~\ref{tab:sig_family_summary_102400}. Lattice signatures (Falcon, ML-DSA) perform well overall. Hash-based SPHINCS+ stands out for its fast verification but slow signing and large signature size. Multivariate MAYO has competitive verification speed and small signatures but suffers from slow signing. RSDP offers fast key generation but slow signing/verification and very large signatures.

\begin{table*}[!ht]
\centering
\caption{Signature family summary statistics (\SI{100}{\kilo\byte} message, macOS)}
\label{tab:sig_family_summary_102400}
\resizebox{\linewidth}{!}{%
\begin{tabular}{@{}l*{20}{r}@{}}
\toprule
& \multicolumn{4}{c}{Key‑generation (ms)} & \multicolumn{4}{c}{Signing (ms)}
& \multicolumn{4}{c}{Verification (ms)} & \multicolumn{4}{c}{Public‑key size (bytes)}
& \multicolumn{4}{c@{}}{Signature size (bytes)}\\
\cmidrule(lr){2-5}\cmidrule(lr){6-9}\cmidrule(lr){10-13}\cmidrule(lr){14-17}\cmidrule(lr){18-21}
Family & mean & std & min & max & mean & std & min & max
& mean & std & min & max & mean & std & min & max
& mean & std & min & max\\
\midrule
ECDSA & 0.14 & 0.11 & 0.01 & 0.22 & 0.27 & 0.18 & 0.06 & 0.40
& 0.46 & 0.32 & 0.08 & 0.66 & 123.00 & 33.60 & 91 & 158
& 104.33 & 33.02 & 72 & 138\\
EdDSA & 0.07 & - & 0.07 & 0.07 & 0.18 & - & 0.18 & 0.18
& 0.21 & - & 0.21 & 0.21 & 44.00 & - & 44 & 44
& 64.00 & - & 64 & 64\\
Falcon & 6.79 & 4.97 & 3.28 & 10.31 & 0.27 & 0.07 & 0.22 & 0.32
& 0.13 & 0.01 & 0.12 & 0.14 & 1345.00 & 633.57 & 897 & 1793
& 963.50 & 436.28 & 655 & 1272\\
Falcon-padded & 6.76 & 4.87 & 3.32 & 10.20 & 0.28 & 0.06 & 0.24 & 0.32
& 0.13 & 0.01 & 0.12 & 0.14 & 1345.00 & 633.57 & 897 & 1793
& 973.00 & 434.16 & 666 & 1280\\
Hash‑based & 29.58 & 31.50 & 0.64 & 88.63 & 293.90 & 293.88 & 15.45 & 799.04
& 1.16 & 0.54 & 0.50 & 2.07 & 48.00 & 13.64 & 32 & 64
& 26080.00 & 14653.83 & 7856 & 49856\\
ML‑DSA & 0.07 & 0.03 & 0.03 & 0.10 & 0.34 & 0.08 & 0.26 & 0.41
& 0.17 & 0.03 & 0.14 & 0.20 & 1952.00 & 640.00 & 1312 & 2592
& 3452.00 & 1110.43 & 2420 & 4627\\
Multivariate & 0.24 & 0.23 & 0.05 & 0.57 & 0.73 & 0.62 & 0.27 & 1.61
& 0.22 & 0.12 & 0.13 & 0.38 & 3580.00 & 2029.05 & 1168 & 5488
& 479.00 & 290.32 & 180 & 838\\
RSA & 209.79 & 182.93 & 54.97 & 411.65 & 4.30 & 2.68 & 1.83 & 7.16
& 0.14 & 0.05 & 0.09 & 0.20 & 422.00 & 128.00 & 294 & 550
& 384.00 & 128.00 & 256 & 512\\
RSDP & 0.02 & 0.01 & 0.00 & 0.04 & 1.62 & 1.20 & 0.31 & 4.71
& 0.94 & 0.62 & 0.22 & 2.35 & 98.00 & 32.49 & 54 & 153
& 29338.11 & 18243.27 & 7956 & 76298\\
\bottomrule
\end{tabular}}%
\end{table*}

\subsection{Deployment Recommendations for Consumer Electronics} \label{sec:deployment_recs}
Synthesizing these findings, we propose tailored PQC recommendations for different CE categories in Table~\ref{tab:deployment_recommendations}.

\begin{table*}[!ht]
\raggedright
\caption{Recommended post‑quantum algorithms for common consumer‑electronics classes}
\label{tab:deployment_recommendations}
\footnotesize
\begin{tabularx}{\textwidth}{@{}>{\bfseries}p{4cm} X X@{}}
\toprule
Class (typical devices) &
\raggedright Recommended KEM(s) &
\raggedright Recommended signature(s) \tabularnewline
\midrule
Low‑power / highly constrained
(\textit{wearables, smart locks, BLE gadgets, MCUs}) &
\textbf{ML‑KEM‑512} (primary); NTRU‑Prime (sntrup761);
hybrid ECDH‑256 for backward compatibility &
\textbf{Falcon‑512} (primary); \textbf{ML‑DSA‑44} (if RAM allows);
Ed25519 (classical until full migration) \tabularnewline[0.3em]
Moderately constrained
(\textit{smart speakers, TVs, home gateways}) &
\textbf{ML‑KEM‑512/768} (primary); BIKE‑L1/L3 or HQC‑128/192;
ECDH‑256/384 for hybrid roll‑outs &
\textbf{ML‑DSA‑44/65}; \textbf{Falcon‑512/1024}; \textbf{SPHINCS+} \textit{f};
MAYO; Ed25519 / ECDSA‑256/384 \tabularnewline[0.3em]
High‑performance
(\textit{smartphones, laptops, game consoles}) &
\textbf{ML‑KEM‑768/1024} (primary); BIKE‑L3/L5, HQC‑192/256;
FrodoKEM; Classic McEliece; ECDH‑384/521 (hybrid) &
\textbf{ML‑DSA‑65/87}; \textbf{Falcon‑1024}; \textbf{SPHINCS+};
MAYO; Ed25519; ECDSA‑384/521; RSDP \tabularnewline
\bottomrule
\end{tabularx}
\vspace{0.4em}
\footnotesize
\textbf{Notes.} Bold items (ML‑KEM, ML‑DSA, Falcon, SPHINCS+) are NIST
standards/finalists. 
Security‑level mapping: L1 $\approx$ 512/44, 
L3 $\approx$ 768/65, 
L5 $\approx$ 1024/87. 
Classical curves are listed only for transitional hybrid use.
\end{table*}

Key takeaways specifically for CE:
\begin{itemize}
    \item \textbf{Prioritize Lattice:} For most common CE scenarios needing a balance of speed, size, and security, the NIST-standardized lattice schemes (ML-KEM, ML-DSA) and finalists (Falcon) are currently the most practical PQC choices.
    \item \textbf{Constrained Bottlenecks:} For very low-end devices, execution time (especially on ARM Cortex-A or M series) and communication/storage size are the primary PQC adoption hurdles. ML-KEM-512 and Falcon-512 offer the lowest overhead among robust PQC options. NTRU-Prime is also efficient.
    \item \textbf{Firmware Signing Choice:} For verifying large firmware OTA updates where verification speed and security assurance are key, but signing is done offline and signature size is less critical than bandwidth during download, Hash-based (SPHINCS+, especially 'f' variants) is a strong contender despite its signing slowness. Falcon is also very good due to fast verification and smaller signatures. ML-DSA offers a good balance.
    \item \textbf{Avoid High Overheads Where Possible:} Algorithms like Code-based KEMs (Classic McEliece key size) and FrodoKEM (software speed/size) present significant integration challenges for typical CE unless specific hardware support or system architectures mitigate these issues. Large signature schemes (Hash-based, RSDP) may be unsuitable for bandwidth-constrained applications.
\end{itemize}
The choice ultimately depends on a detailed analysis of the specific CE device's resources, use case requirements (latency sensitivity, data volume, update frequency), and security lifetime needs.


\section{Conclusion}
\label{sec:conclusion}
The impending threat of quantum computing necessitates a proactive transition to post-quantum cryptography (PQC) within the vast consumer electronics (CE) ecosystem. This paper addressed this challenge by delivering a comprehensive performance analysis of leading PQC algorithms across platforms representative of diverse CE device capabilities, from high-performance systems to resource-constrained proxies like the Raspberry Pi 4.

Our quantitative evaluation confirmed that while PQC integration is feasible, it demands careful consideration of performance trade-offs, which are significantly amplified on resource-limited hardware typical in CE. The NIST-standardized lattice-based schemes ML-KEM and ML-DSA, along with Falcon, emerged as particularly compelling choices. They offer a practical balance between computational efficiency and manageable communication/storage overheads, crucial attributes for deployment in a wide range of CE applications. Conversely, the substantial key sizes of schemes like Classic McEliece and the large signatures generated by hash-based SPHINCS+ present tangible integration hurdles for bandwidth-sensitive or storage-constrained CE devices, despite their respective security merits. SPHINCS+ remains a strong option for specific use cases like secure firmware verification where its slow signing and large signature size are less critical than its high security assurance and verification speed.

The primary contribution of this work lies in providing empirical, cross-platform performance data and derived, specific deployment recommendations (summarized in Table~\ref{tab:deployment_recommendations}) tailored to different CE classes. These actionable insights directly aid manufacturers, designers, and developers in selecting appropriate PQC solutions based on concrete metrics aligned with device constraints, security requirements, and application contexts. By quantifying these critical performance characteristics and trade-offs, this study provides a vital foundation for navigating the complex migration towards quantum-resistant security in consumer electronics, ensuring the long-term trustworthiness of connected devices.

\section{Future Work}
\label{sec:future_work}
Building upon the analysis presented in this paper, several avenues for future research are pertinent to advancing the practical deployment of PQC in consumer electronics:

\begin{enumerate}
    \item \textbf{Microcontroller Benchmarking:} Extending performance evaluations to true microcontroller platforms (e.g., ARM Cortex-M series) prevalent in low-power wearables and IoT endpoints to assess feasibility and resource usage in highly constrained environments.
    \item \textbf{Hardware Acceleration for CE:} Investigating, designing, and evaluating dedicated hardware accelerators for PQC primitives (particularly lattice-based operations) suitable for cost-effective integration into CE System-on-Chips (SoCs) to mitigate performance overheads \cite{banerjee_accelerating_2020}.
    \item \textbf{Energy Consumption Analysis:} Conducting direct measurements of energy consumption for PQC operations on representative battery-powered CE hardware, moving beyond execution time as a proxy to accurately model battery life impact \cite{roma_energy_2021}.
    \item \textbf{Hybrid Schemes in CE Contexts:} Analyzing the performance, overhead (code size, memory, latency), and transitional complexities of deploying hybrid PQC/classical cryptographic schemes specifically within CE systems and protocols \cite{joseph_transitioning_2022}.
    \item \textbf{Side-Channel Analysis and Mitigation for CE:} Evaluating the susceptibility of PQC implementations to side-channel attacks (timing, power) on typical CE processors and developing efficient, low-overhead countermeasures appropriate for constrained devices \cite{xin_vpqc_2020}.
    \item \textbf{Real-world CE Protocol Integration:} Benchmarking PQC algorithms integrated into actual CE communication protocols (e.g., TLS 1.3 handshakes, Matter secure channel establishment, Bluetooth pairing) on representative hardware to understand end-to-end performance impacts and integration challenges.
    \item \textbf{Memory Footprint Investigation:} Developing and applying reliable, cross-platform methodologies to accurately measure and compare the static and dynamic RAM footprint of different PQC implementations during cryptographic operations on CE-relevant hardware and operating systems.
\end{enumerate}
Addressing these areas will further facilitate the secure and efficient adoption of post-quantum cryptography across the diverse landscape of consumer electronics.


\bibliographystyle{IEEEtran}
\bibliography{tce} 

\begin{thebibliography}{10}
\providecommand{\url}[1]{#1}
\csname url@samestyle\endcsname
\providecommand{\newblock}{\relax}
\providecommand{\bibinfo}[2]{#2}
\providecommand{\BIBentrySTDinterwordspacing}{\spaceskip=0pt\relax}
\providecommand{\BIBentryALTinterwordstretchfactor}{4}
\providecommand{\BIBentryALTinterwordspacing}{\spaceskip=\fontdimen2\font plus
\BIBentryALTinterwordstretchfactor\fontdimen3\font minus \fontdimen4\font\relax}
\providecommand{\BIBforeignlanguage}[2]{{%
\expandafter\ifx\csname l@#1\endcsname\relax
\typeout{** WARNING: IEEEtran.bst: No hyphenation pattern has been}%
\typeout{** loaded for the language `#1'. Using the pattern for}%
\typeout{** the default language instead.}%
\else
\language=\csname l@#1\endcsname
\fi
#2}}
\providecommand{\BIBdecl}{\relax}
\BIBdecl

\bibitem{keranen_cryptographic_2014}
\BIBentryALTinterwordspacing
V.~Keränen, ``Cryptographic algorithm benchmarking in mobile devices,'' Master's thesis, V. Keränen, 2014. [Online]. Available: \url{https://oulurepo.oulu.fi/handle/10024/39404}
\BIBentrySTDinterwordspacing

\bibitem{bui_health_2011}
N.~Bui and M.~Zorzi, ``\BIBforeignlanguage{en}{Health care applications: a solution based on the internet of things},'' in \emph{\BIBforeignlanguage{en}{Proceedings of the 4th {International} {Symposium} on {Applied} {Sciences} in {Biomedical} and {Communication} {Technologies}}}.\hskip 1em plus 0.5em minus 0.4em\relax Barcelona Spain: ACM, Oct. 2011, pp. 1--5.

\bibitem{stallings_cryptography_2017}
W.~Stallings, \emph{Cryptography \& {Network} {Security} {GE}.}\hskip 1em plus 0.5em minus 0.4em\relax Pearson Australia Pty Limited, 2017.

\bibitem{shor_polynomial-time_1999}
P.~W. Shor, ``Polynomial-time algorithms for prime factorization and discrete logarithms on a quantum computer,'' \emph{SIAM review}, vol.~41, no.~2, pp. 303--332, 1999.

\bibitem{mosca_quantum_2021}
\BIBentryALTinterwordspacing
M.~Mosca and M.~Piani, ``Quantum threat timeline report 2020,'' \emph{Global Risk Insitute: https://globalriskinstitute. org/publications/quantum-threat-timeline-report-2020}, 2021. [Online]. Available: \url{https://quantum-safe.ca/wp-content/uploads/2023/01/2022-quantum-threat-timeline-report-dec.pdf}
\BIBentrySTDinterwordspacing

\bibitem{grumbling_quantum_2019}
\BIBentryALTinterwordspacing
E.~Grumbling and M.~Horowitz, Eds., \emph{Quantum {Computing}: {Progress} and {Prospects}}.\hskip 1em plus 0.5em minus 0.4em\relax Washington, D.C.: National Academies Press, Mar. 2019. [Online]. Available: \url{https://www.nap.edu/catalog/25196}
\BIBentrySTDinterwordspacing

\bibitem{bernstein_post-quantum_2017}
D.~Bernstein and T.~Lange, ``Post-quantum cryptography,'' \emph{NATURE}, vol. 549, no. 7671, pp. 188--194, Sep. 2017.

\bibitem{alagic_status_2022}
\BIBentryALTinterwordspacing
G.~Alagic, D.~Apon, D.~Cooper, Q.~Dang, T.~Dang, J.~Kelsey, J.~Lichtinger, C.~Miller, D.~Moody, R.~Peralta, R.~Perlner, A.~Robinson, D.~Smith-Tone, and Y.-K. Liu, ``\BIBforeignlanguage{en}{Status {Report} on the {Third} {Round} of the {NIST} {Post}-{Quantum} {Cryptography} {Standardization} {Process}},'' National Institute of Standards and Technology, Tech. Rep. NIST Internal or Interagency Report (NISTIR) 8413, Sep. 2022. [Online]. Available: \url{https://csrc.nist.gov/pubs/ir/8413/upd1/final}
\BIBentrySTDinterwordspacing

\bibitem{NIST_FIPS_204}
\BIBentryALTinterwordspacing
{National Institute of Standards and Technology (US)}, ``\BIBforeignlanguage{en}{Module-lattice-based digital signature standard},'' National Institute of Standards and Technology (U.S.), Washington, D.C., Tech. Rep. NIST FIPS 204, Aug. 2024. [Online]. Available: \url{https://nvlpubs.nist.gov/nistpubs/FIPS/NIST.FIPS.204.pdf}
\BIBentrySTDinterwordspacing

\bibitem{NIST_FIPS_205}
\BIBentryALTinterwordspacing
------, ``\BIBforeignlanguage{en}{Stateless hash-based digital signature standard},'' National Institute of Standards and Technology (U.S.), Washington, D.C., Tech. Rep. NIST FIPS 205, Aug. 2024. [Online]. Available: \url{https://nvlpubs.nist.gov/nistpubs/FIPS/NIST.FIPS.205.pdf}
\BIBentrySTDinterwordspacing

\bibitem{paquin_benchmarking_2020}
C.~Paquin, D.~Stebila, and G.~Tamvada, ``\BIBforeignlanguage{en}{Benchmarking {Post}-quantum {Cryptography} in {TLS}},'' in \emph{\BIBforeignlanguage{en}{Post-{Quantum} {Cryptography}}}, J.~Ding and J.-P. Tillich, Eds.\hskip 1em plus 0.5em minus 0.4em\relax Cham: Springer International Publishing, 2020, vol. 12100, pp. 72--91.

\bibitem{sikeridis_post-quantum_2020}
\BIBentryALTinterwordspacing
D.~Sikeridis, P.~Kampanakis, and M.~Devetsikiotis, ``Post-quantum authentication in {TLS} 1.3: a performance study,'' \emph{Cryptology ePrint Archive}, 2020. [Online]. Available: \url{https://eprint.iacr.org/2020/071}
\BIBentrySTDinterwordspacing

\bibitem{basu_nist_2019}
\BIBentryALTinterwordspacing
K.~Basu, D.~Soni, M.~Nabeel, and R.~Karri, ``Nist post-quantum cryptography-a hardware evaluation study,'' \emph{Cryptology ePrint Archive}, 2019. [Online]. Available: \url{https://eprint.iacr.org/2019/047}
\BIBentrySTDinterwordspacing

\bibitem{raavi_performance_2021}
\BIBentryALTinterwordspacing
M.~Raavi, P.~Chandramouli, S.~Wuthier, X.~Zhou, and S.-Y. Chang, ``Performance characterization of post-quantum digital certificates,'' in \emph{2021 {International} {Conference} on {Computer} {Communications} and {Networks} ({ICCCN})}.\hskip 1em plus 0.5em minus 0.4em\relax IEEE, 2021, pp. 1--9. [Online]. Available: \url{https://ieeexplore.ieee.org/abstract/document/9522179/}
\BIBentrySTDinterwordspacing

\bibitem{septien-hernandez_comparative_2022}
\BIBentryALTinterwordspacing
J.-A. Septien-Hernandez, M.~Arellano-Vazquez, M.~A. Contreras-Cruz, and J.-P. Ramirez-Paredes, ``A {Comparative} study of post-quantum cryptosystems for {Internet}-of-{Things} applications,'' \emph{Sensors}, vol.~22, no.~2, p. 489, 2022. [Online]. Available: \url{https://www.mdpi.com/1424-8220/22/2/489}
\BIBentrySTDinterwordspacing

\bibitem{kumar_securing_2022}
A.~Kumar, C.~Ottaviani, S.~S. Gill, and R.~Buyya, ``\BIBforeignlanguage{en}{Securing the future internet of things with post‐quantum cryptography},'' \emph{\BIBforeignlanguage{en}{SECURITY AND PRIVACY}}, vol.~5, no.~2, p. e200, Mar. 2022, \_eprint: 2206.10473.

\bibitem{bernstein_analyzing_2023}
\BIBentryALTinterwordspacing
D.~J. Bernstein, ``Analyzing the complexity of reference post-quantum software: the case of lattice-based {KEMs},'' \emph{Cryptology ePrint Archive}, 2023. [Online]. Available: \url{https://eprint.iacr.org/2023/1924}
\BIBentrySTDinterwordspacing

\bibitem{satrya_comparative_2023}
\BIBentryALTinterwordspacing
G.~B. Satrya, Y.~M. Agus, and A.~B. Mnaouer, ``A comparative study of post-quantum cryptographic algorithm implementations for secure and efficient energy systems monitoring,'' \emph{Electronics}, vol.~12, no.~18, p. 3824, 2023. [Online]. Available: \url{https://www.mdpi.com/2079-9292/12/18/3824}
\BIBentrySTDinterwordspacing

\bibitem{prantl_performance_2021}
\BIBentryALTinterwordspacing
T.~Prantl, D.~Prantl, L.~Beierlieb, L.~Iffländer, A.~Dmitrienko, S.~Kounev, and C.~Krupitzer, ``Performance {Evaluation} for a {Post}-{Quantum} {Public}-{Key} {Cryptosystem},'' in \emph{2021 {IEEE} {International} {Performance}, {Computing}, and {Communications} {Conference} ({IPCCC})}.\hskip 1em plus 0.5em minus 0.4em\relax IEEE, 2021, pp. 1--7. [Online]. Available: \url{https://ieeexplore.ieee.org/abstract/document/9679412/}
\BIBentrySTDinterwordspacing

\bibitem{lakhan_comparative_2023}
\BIBentryALTinterwordspacing
A.~S. Lakhan, ``A {Comparative} {Study} on {Post}-{Quantum} {Cryptographic} {Digital} {Signature} {Algorithms}: {Network} {Performance}, {Key} {Robustness}, and {Energy} {Consumption}.'' {PhD} {Thesis}, Carleton University, 2023. [Online]. Available: \url{https://repository.library.carleton.ca/concern/etds/3197xn336}
\BIBentrySTDinterwordspacing

\bibitem{xin_vpqc_2020}
\BIBentryALTinterwordspacing
G.~Xin, J.~Han, T.~Yin, Y.~Zhou, J.~Yang, X.~Cheng, and X.~Zeng, ``{VPQC}: {A} domain-specific vector processor for post-quantum cryptography based on {RISC}-{V} architecture,'' \emph{IEEE transactions on circuits and systems I: regular papers}, vol.~67, no.~8, pp. 2672--2684, 2020. [Online]. Available: \url{https://ieeexplore.ieee.org/abstract/document/9061149/}
\BIBentrySTDinterwordspacing

\bibitem{fritzmann_risq-v_2020}
\BIBentryALTinterwordspacing
T.~Fritzmann, G.~Sigl, and J.~Sepúlveda, ``{RISQ}-{V}: {Tightly} coupled {RISC}-{V} accelerators for post-quantum cryptography,'' \emph{IACR Transactions on Cryptographic Hardware and Embedded Systems}, pp. 239--280, 2020. [Online]. Available: \url{https://tches.iacr.org/index.php/TCHES/article/view/8683}
\BIBentrySTDinterwordspacing

\bibitem{pomerance_tale_1996}
C.~Pomerance, ``A tale of two sieves,'' \emph{Notices of the American Mathematical Society}, vol.~43, no.~12, pp. 1473--1485, 1996.

\bibitem{nielsen_quantum_2010}
M.~A. Nielsen and I.~L. Chuang, \emph{Quantum computation and quantum information}.\hskip 1em plus 0.5em minus 0.4em\relax Cambridge university press, 2010.

\bibitem{grover_fast_1996}
L.~Grover, ``\BIBforeignlanguage{en}{A fast quantum mechanical algorithm for database search},'' in \emph{\BIBforeignlanguage{en}{Proceedings of the twenty-eighth annual {ACM} symposium on {Theory} of computing - {STOC} ’96}}.\hskip 1em plus 0.5em minus 0.4em\relax ACM Press, 1996, pp. 212--219.

\bibitem{mina_information_2021}
\BIBentryALTinterwordspacing
M.-Z. Mina and E.~Simion, ``Information {Security} in the {Quantum} {Era}. {Threats} to modern cryptography: {Grover}’s algorithm,'' \emph{Cryptology ePrint Archive}, 2021. [Online]. Available: \url{https://eprint.iacr.org/2021/1662}
\BIBentrySTDinterwordspacing

\bibitem{regev_lattices_2009}
O.~Regev, ``\BIBforeignlanguage{en}{On lattices, learning with errors, random linear codes, and cryptography},'' \emph{\BIBforeignlanguage{en}{Journal of the ACM}}, vol.~56, no.~6, pp. 1--40, Sep. 2009.

\bibitem{lyubashevsky_toolkit_2013}
V.~Lyubashevsky, C.~Peikert, and O.~Regev, ``\BIBforeignlanguage{en}{A {Toolkit} for {Ring}-{LWE} {Cryptography}},'' in \emph{\BIBforeignlanguage{en}{Advances in {Cryptology} – {EUROCRYPT} 2013}}, D.~Hutchison, T.~Kanade, J.~Kittler, T.~Johansson, J.~M. Kleinberg, F.~Mattern, J.~C. Mitchell, M.~Naor, O.~Nierstrasz, C.~Pandu~Rangan, B.~Steffen, M.~Sudan, D.~Terzopoulos, D.~Tygar, M.~Y. Vardi, G.~Weikum, and P.~Q. Nguyen, Eds.\hskip 1em plus 0.5em minus 0.4em\relax Berlin, Heidelberg: Springer Berlin Heidelberg, 2013, vol. 7881, pp. 35--54.

\bibitem{bos_crystals-kyber_2018}
\BIBentryALTinterwordspacing
J.~Bos, L.~Ducas, E.~Kiltz, T.~Lepoint, V.~Lyubashevsky, J.~M. Schanck, P.~Schwabe, G.~Seiler, and D.~Stehlé, ``{CRYSTALS}-{Kyber}: a {CCA}-secure module-lattice-based {KEM},'' in \emph{2018 {IEEE} {European} {Symposium} on {Security} and {Privacy} ({EuroS}\&{P})}.\hskip 1em plus 0.5em minus 0.4em\relax IEEE, 2018, pp. 353--367. [Online]. Available: \url{https://ieeexplore.ieee.org/abstract/document/8406610/}
\BIBentrySTDinterwordspacing

\bibitem{ducas_crystals-dilithium_2018}
\BIBentryALTinterwordspacing
L.~Ducas, E.~Kiltz, T.~Lepoint, V.~Lyubashevsky, P.~Schwabe, G.~Seiler, and D.~Stehlé, ``Crystals-dilithium: {A} lattice-based digital signature scheme,'' \emph{IACR Transactions on Cryptographic Hardware and Embedded Systems}, pp. 238--268, 2018. [Online]. Available: \url{https://tches.iacr.org/index.php/TCHES/article/view/839}
\BIBentrySTDinterwordspacing

\bibitem{fouque_falcon_2018}
\BIBentryALTinterwordspacing
P.-A. Fouque, J.~Hoffstein, P.~Kirchner, V.~Lyubashevsky, T.~Pornin, T.~Prest, T.~Ricosset, G.~Seiler, W.~Whyte, and Z.~Zhang, ``Falcon: {Fast}-{Fourier} lattice-based compact signatures over {NTRU},'' \emph{Submission to the NIST’s post-quantum cryptography standardization process}, vol.~36, no.~5, pp. 1--75, 2018. [Online]. Available: \url{https://www.di.ens.fr/~prest/Publications/falcon.pdf}
\BIBentrySTDinterwordspacing

\bibitem{bernstein_ntru_2016}
\BIBentryALTinterwordspacing
D.~J. Bernstein, C.~Chuengsatiansup, T.~Lange, and C.~Van~Vredendaal, ``{NTRU} {Prime}.'' \emph{IACR Cryptol. ePrint Arch.}, vol. 2016, p. 461, 2016. [Online]. Available: \url{http://hyperelliptic.org/tanja/vortraege/caen-ntruprime.pdf}
\BIBentrySTDinterwordspacing

\bibitem{berlekamp_inherent_2003}
\BIBentryALTinterwordspacing
E.~Berlekamp, R.~McEliece, and H.~Van~Tilborg, ``On the inherent intractability of certain coding problems (corresp.),'' \emph{IEEE Transactions on Information theory}, vol.~24, no.~3, pp. 384--386, 2003. [Online]. Available: \url{https://ieeexplore.ieee.org/abstract/document/1055873/}
\BIBentrySTDinterwordspacing

\bibitem{mceliece_public-key_1978}
\BIBentryALTinterwordspacing
R.~J. McEliece, ``A public-key cryptosystem based on algebraic,'' \emph{Coding Thv}, vol. 4244, no. 1978, pp. 114--116, 1978. [Online]. Available: \url{https://ntrs.nasa.gov/api/citations/19780016269/downloads/19780016269.pdf#page=123}
\BIBentrySTDinterwordspacing

\bibitem{baldi_post-quantum_2017}
\BIBentryALTinterwordspacing
M.~Baldi, P.~Santini, and G.~Cancellieri, ``Post-quantum cryptography based on codes: {State} of the art and open challenges,'' in \emph{2017 {AEIT} {International} {Annual} {Conference}}.\hskip 1em plus 0.5em minus 0.4em\relax IEEE, 2017, pp. 1--6. [Online]. Available: \url{https://ieeexplore.ieee.org/abstract/document/8240549/}
\BIBentrySTDinterwordspacing

\bibitem{aragon_bike_2022}
\BIBentryALTinterwordspacing
N.~Aragon, P.~Barreto, S.~Bettaieb, L.~Bidoux, O.~Blazy, J.-C. Deneuville, P.~Gaborit, S.~Ghosh, S.~Gueron, and T.~Güneysu, ``{BIKE}: bit flipping key encapsulation,'' 2022. [Online]. Available: \url{https://inria.hal.science/hal-04278509/document}
\BIBentrySTDinterwordspacing

\bibitem{melchor_hamming_2018}
\BIBentryALTinterwordspacing
C.~A. Melchor, N.~Aragon, S.~Bettaieb, L.~Bidoux, O.~Blazy, J.-C. Deneuville, P.~Gaborit, E.~Persichetti, G.~Zémor, and I.~C. Bourges, ``Hamming quasi-cyclic ({HQC}),'' \emph{NIST PQC Round}, vol.~2, no.~4, p.~13, 2018. [Online]. Available: \url{https://pqc-hqc.org/doc/hqc-specification_2023-04-30.pdf}
\BIBentrySTDinterwordspacing

\bibitem{lamport_constructing_1979}
\BIBentryALTinterwordspacing
L.~Lamport, ``Constructing digital signatures from a one way function,'' 1979. [Online]. Available: \url{https://www.microsoft.com/en-us/research/publication/constructing-digital-signatures-one-way-function/}
\BIBentrySTDinterwordspacing

\bibitem{buchmann_security_2011}
J.~Buchmann, E.~Dahmen, A.~Hülsing, S.~Ereth, and M.~Rückert, ``On the {Security} of the {Winternitz} {One}-{Time} {Signature} {Scheme},'' in \emph{Progress in {Cryptology} – {AFRICACRYPT} 2011}, A.~Nitaj and D.~Pointcheval, Eds.\hskip 1em plus 0.5em minus 0.4em\relax Berlin, Heidelberg: Springer Berlin Heidelberg, 2011, vol. 6737, pp. 363--378.

\bibitem{hulsing_xmss_2018}
A.~T. Hülsing, D.~Butin, S.-L. Gazdag, J.~Rijneveld, and A.~Mohaisen, ``{XMSS}: extended hash-based signatures. {RFC} 8391,'' \emph{Request for Comments}, 2018.

\bibitem{bernstein_sphincs_2019}
D.~J. Bernstein, A.~Hülsing, S.~Kölbl, R.~Niederhagen, J.~Rijneveld, and P.~Schwabe, ``\BIBforeignlanguage{en}{The {SPHINCS}\textsuperscript{+} {Signature} framework},'' in \emph{\BIBforeignlanguage{en}{Proceedings of the 2019 {ACM} {SIGSAC} {Conference} on {Computer} and {Communications} {Security}}}.\hskip 1em plus 0.5em minus 0.4em\relax London United Kingdom: ACM, Nov. 2019, pp. 2129--2146.

\bibitem{kudinov_sphincs_2022}
\BIBentryALTinterwordspacing
M.~Kudinov, A.~Hülsing, E.~Ronen, and E.~Yogev, ``{SPHINCS}+ {C}: {Compressing} {SPHINCS}+ with (almost) no cost,'' \emph{Cryptology ePrint Archive}, 2022. [Online]. Available: \url{https://eprint.iacr.org/2022/778}
\BIBentrySTDinterwordspacing

\bibitem{garey_computers_2002}
\BIBentryALTinterwordspacing
M.~R. Garey and D.~S. Johnson, \emph{Computers and intractability}.\hskip 1em plus 0.5em minus 0.4em\relax wh freeman New York, 2002, vol.~29. [Online]. Available: \url{https://bohr.wlu.ca/hfan/cp412/references/ChapterOne.pdf}
\BIBentrySTDinterwordspacing

\bibitem{ding_multivariate_2006}
J.~Ding, J.~E. Gower, and D.~S. Schmidt, \emph{Multivariate public key cryptosystems}.\hskip 1em plus 0.5em minus 0.4em\relax Springer Science \& Business Media, 2006, vol.~25.

\bibitem{beullens_breaking_2022}
W.~Beullens, ``\BIBforeignlanguage{en}{Breaking {Rainbow} {Takes} a {Weekend} on a {Laptop}},'' in \emph{\BIBforeignlanguage{en}{Advances in {Cryptology} – {CRYPTO} 2022}}, Y.~Dodis and T.~Shrimpton, Eds.\hskip 1em plus 0.5em minus 0.4em\relax Cham: Springer Nature Switzerland, 2022, vol. 13508, pp. 464--479.

\bibitem{beullens_mayo_2022}
------, ``\BIBforeignlanguage{en}{{MAYO}: {Practical} {Post}-quantum {Signatures} from {Oil}-and-{Vinegar} {Maps}},'' in \emph{\BIBforeignlanguage{en}{Selected {Areas} in {Cryptography}}}, R.~AlTawy and A.~Hülsing, Eds.\hskip 1em plus 0.5em minus 0.4em\relax Cham: Springer International Publishing, 2022, vol. 13203, pp. 355--376.

\bibitem{jao_sike_2017}
\BIBentryALTinterwordspacing
D.~Jao, R.~Azarderakhsh, M.~Campagna, C.~Costello, L.~De~Feo, B.~Hess, A.~Jalili, B.~Koziel, B.~LaMacchia, and P.~Longa, ``{SIKE}: {Supersingular} isogeny key encapsulation,'' 2017. [Online]. Available: \url{https://hal.science/hal-02171951/}
\BIBentrySTDinterwordspacing

\bibitem{castryck_efficient_2023}
W.~Castryck and T.~Decru, ``\BIBforeignlanguage{en}{An {Efficient} {Key} {Recovery} {Attack} on {SIDH}},'' in \emph{\BIBforeignlanguage{en}{Advances in {Cryptology} – {EUROCRYPT} 2023}}, C.~Hazay and M.~Stam, Eds.\hskip 1em plus 0.5em minus 0.4em\relax Cham: Springer Nature Switzerland, 2023, vol. 14008, pp. 423--447.

\bibitem{singh_code_2020}
H.~Singh, ``Code based {Cryptography}: {Classic} {McEliece},'' May 2020, arXiv: 1907.12754.

\bibitem{ding_new_2004}
J.~Ding, ``A {New} {Variant} of the {Matsumoto}-{Imai} {Cryptosystem} through {Perturbation},'' in \emph{Public {Key} {Cryptography} – {PKC} 2004}, G.~Goos, J.~Hartmanis, J.~Van~Leeuwen, F.~Bao, R.~Deng, and J.~Zhou, Eds.\hskip 1em plus 0.5em minus 0.4em\relax Berlin, Heidelberg: Springer Berlin Heidelberg, 2004, vol. 2947, pp. 305--318.

\bibitem{aydin_horizontal_2021}
F.~Aydin, A.~Aysu, M.~Tiwari, A.~Gerstlauer, and M.~Orshansky, ``\BIBforeignlanguage{en}{Horizontal {Side}-{Channel} {Vulnerabilities} of {Post}-{Quantum} {Key} {Exchange} and {Encapsulation} {Protocols}},'' \emph{\BIBforeignlanguage{en}{ACM Transactions on Embedded Computing Systems}}, vol.~20, no.~6, pp. 1--22, Nov. 2021.

\bibitem{stebila_post-quantum_2017}
D.~Stebila and M.~Mosca, ``\BIBforeignlanguage{en}{Post-quantum {Key} {Exchange} for the {Internet} and the {Open} {Quantum} {Safe} {Project}},'' in \emph{\BIBforeignlanguage{en}{Selected {Areas} in {Cryptography} – {SAC} 2016}}, R.~Avanzi and H.~Heys, Eds.\hskip 1em plus 0.5em minus 0.4em\relax Cham: Springer International Publishing, 2017, vol. 10532, pp. 14--37.

\bibitem{banerjee_accelerating_2020}
\BIBentryALTinterwordspacing
U.~Banerjee, S.~Das, and A.~P. Chandrakasan, ``Accelerating post-quantum cryptography using an energy-efficient {TLS} crypto-processor,'' in \emph{2020 {IEEE} {International} {Symposium} on {Circuits} and {Systems} ({ISCAS})}.\hskip 1em plus 0.5em minus 0.4em\relax IEEE, 2020, pp. 1--5. [Online]. Available: \url{https://ieeexplore.ieee.org/abstract/document/9180550/}
\BIBentrySTDinterwordspacing

\bibitem{roma_energy_2021}
\BIBentryALTinterwordspacing
C.~A. Roma, C.-E.~A. Tai, and M.~A. Hasan, ``Energy efficiency analysis of post-quantum cryptographic algorithms,'' \emph{IEEE Access}, vol.~9, pp. 71\,295--71\,317, 2021. [Online]. Available: \url{https://ieeexplore.ieee.org/abstract/document/9424003/}
\BIBentrySTDinterwordspacing

\bibitem{joseph_transitioning_2022}
\BIBentryALTinterwordspacing
D.~Joseph, R.~Misoczki, M.~Manzano, J.~Tricot, F.~D. Pinuaga, O.~Lacombe, S.~Leichenauer, J.~Hidary, P.~Venables, and R.~Hansen, ``Transitioning organizations to post-quantum cryptography,'' \emph{Nature}, vol. 605, no. 7909, pp. 237--243, 2022. [Online]. Available: \url{https://idp.nature.com/authorize/casa?redirect_uri=https://www.nature.com/articles/s41586-022-04623-2}
\BIBentrySTDinterwordspacing

\end{thebibliography}

\onecolumn
\appendices

\section{Detailed Performance and Ratio Analysis}
\label{sec:appendix_details}

This appendix provides supplementary data supporting the analysis presented in the main body of the paper. The goal is to offer a granular view of individual algorithm performance and quantify the performance differences observed across the evaluated hardware platforms. It includes detailed per-algorithm performance metrics measured on the macOS reference platform, facilitating direct comparisons between specific parameter sets and algorithm variants. Furthermore, it presents platform performance ratio comparisons, highlighting the relative speed differences between the Ubuntu desktop and Raspberry Pi constrained platforms compared to the macOS reference.

\subsection{Detailed Performance Tables (macOS Reference Platform)}

The following tables detail the performance characteristics of each tested cryptographic algorithm on the macOS M4 reference platform. This platform serves as a baseline representing a modern, relatively high-performance consumer device. Understanding performance on this reference platform is crucial for comparing the inherent computational costs of different algorithms before considering the impact of resource constraints.

Table~\ref{tab:appendix_kem_detailed} presents the detailed performance and size metrics for all evaluated Key Encapsulation Mechanisms (KEMs). This includes mean execution times (in milliseconds) for the three core KEM operations: key generation, encapsulation (generating a shared secret and its corresponding ciphertext), and decapsulation (recovering the shared secret from the ciphertext). It also lists the associated communication and storage overheads: the size (in bytes) of the public key required by the encapsulator, the size of the ciphertext transmitted, and the size of the resulting shared secret. These metrics allow for a fine-grained comparison of KEM candidates based on both speed and size for various security levels (NIST security levels 1, 2, 3, 5 where applicable). Note the inclusion of both standard ('f') and alternative Classic McEliece implementations where available in the underlying benchmark suite.

\begin{table*}[!htbp]
\centering
\caption{Detailed KEM performance and size metrics (macOS)}
\label{tab:appendix_kem_detailed}
\resizebox{\linewidth}{!}{%
\begin{tabular}{@{}lllrrrrrrr@{}}
\toprule
Algorithm & Family & Type & Sec.\ Level & Key‑gen (ms) & Encaps.\ (ms) & Decaps.\ (ms) & Pub‑key (B) & Ciphertext (B) & Shared key (B) \\
\midrule
BIKE-L1 & Code-based & PQ & 1 & 4.19 & 0.22 & 3.22 & 1,541 & 1,573 & 32 \\
Classic-McEliece-348864 & Code-based & PQ & 1 & 44.75 & 0.02 & 12.81 & 261,120 & 96 & 32 \\
Classic-McEliece-348864f & Code-based & PQ & 1 & 21.71 & 0.02 & 12.71 & 261,120 & 96 & 32 \\
HQC-128 & Code-based & PQ & 1 & 0.32 & 0.65 & 1.00 & 2,249 & 4,433 & 64 \\
BIKE-L3 & Code-based & PQ & 3 & 12.90 & 0.67 & 10.04 & 3,083 & 3,115 & 32 \\
Classic-McEliece-460896 & Code-based & PQ & 3 & 125.59 & 0.05 & 21.64 & 524,160 & 156 & 32 \\
Classic-McEliece-460896f & Code-based & PQ & 3 & 61.03 & 0.05 & 21.57 & 524,160 & 156 & 32 \\
HQC-192 & Code-based & PQ & 3 & 0.96 & 1.92 & 2.90 & 4,522 & 8,978 & 64 \\
BIKE-L5 & Code-based & PQ & 5 & 32.26 & 1.67 & 25.04 & 5,122 & 5,154 & 32 \\
Classic-McEliece-6688128 & Code-based & PQ & 5 & 264.86 & 0.11 & 41.80 & 1,044,992 & 208 & 32 \\
Classic-McEliece-6688128f & Code-based & PQ & 5 & 99.29 & 0.11 & 41.57 & 1,044,992 & 208 & 32 \\
Classic-McEliece-6960119 & Code-based & PQ & 5 & 249.89 & 0.43 & 40.48 & 1,047,319 & 194 & 32 \\
Classic-McEliece-6960119f & Code-based & PQ & 5 & 98.21 & 0.42 & 40.37 & 1,047,319 & 194 & 32 \\
Classic-McEliece-8192128 & Code-based & PQ & 5 & 285.45 & 0.12 & 51.34 & 1,357,824 & 208 & 32 \\
Classic-McEliece-8192128f & Code-based & PQ & 5 & 105.91 & 0.11 & 51.09 & 1,357,824 & 208 & 32 \\
HQC-256 & Code-based & PQ & 5 & 1.76 & 3.47 & 5.28 & 7,245 & 14,421 & 64 \\
FrodoKEM-640-AES & FrodoKEM & PQ & 1 & 0.16 & 0.25 & 0.24 & 9,616 & 9,720 & 16 \\
FrodoKEM-640-SHAKE & FrodoKEM & PQ & 1 & 0.81 & 1.02 & 0.90 & 9,616 & 9,720 & 16 \\
FrodoKEM-976-AES & FrodoKEM & PQ & 3 & 0.33 & 0.49 & 0.48 & 15,632 & 15,744 & 24 \\
FrodoKEM-976-SHAKE & FrodoKEM & PQ & 3 & 1.81 & 2.04 & 2.02 & 15,632 & 15,744 & 24 \\
FrodoKEM-1344-AES & FrodoKEM & PQ & 5 & 0.57 & 0.85 & 0.83 & 21,520 & 21,632 & 32 \\
FrodoKEM-1344-SHAKE & FrodoKEM & PQ & 5 & 3.25 & 3.68 & 3.66 & 21,520 & 21,632 & 32 \\
ML-KEM-512 & ML-KEM & PQ & 1 & 0.01 & 0.01 & 0.01 & 800 & 768 & 32 \\
ML-KEM-768 & ML-KEM & PQ & 3 & 0.01 & 0.01 & 0.01 & 1,184 & 1,088 & 32 \\
ML-KEM-1024 & ML-KEM & PQ & 5 & 0.01 & 0.01 & 0.02 & 1,568 & 1,568 & 32 \\
sntrup761 & NTRU-Prime & PQ & 2 & 1.40 & 0.04 & 0.07 & 1,158 & 1,039 & 32 \\
ECDH-256 & ECDH & Classical & 1 & 0.02 & 0.05 & 0.05 & 91 & 91 & 32 \\
ECDH-384 & ECDH & Classical & 3 & 0.20 & 0.50 & 0.49 & 120 & 120 & 48 \\
ECDH-521 & ECDH & Classical & 5 & 0.21 & 0.46 & 0.46 & 158 & 158 & 66 \\
RSA-2048 & RSA & Classical & 1 & 56.84 & 0.06 & 1.73 & 294 & 256 & 32 \\
RSA-3072 & RSA & Classical & 3 & 160.74 & 0.11 & 3.84 & 422 & 384 & 32 \\
RSA-4096 & RSA & Classical & 5 & 395.10 & 0.17 & 6.96 & 550 & 512 & 32 \\
\bottomrule
\end{tabular}}%
\end{table*}

Table~\ref{tab:appendix_sig_detailed} provides similar detailed metrics for the evaluated digital signature schemes on the macOS platform. It shows the mean key generation time and the public key size. Crucially, it breaks down performance for the core signature operations (signing and verification) across different message sizes: \SI{1}{\kilo\byte}, \SI{10}{\kilo\byte}, \SI{100}{\kilo\byte}, and \SI{1}{\mega\byte} (\SI{1024}{\kilo\byte}). This allows analysis of how algorithm performance scales with the amount of data being signed or verified, a critical factor for applications like secure boot or firmware updates. The table also includes the resulting signature size for each algorithm and message size combination, although for many schemes (like ML-DSA, Falcon, SPHINCS+), the signature size is independent of the message size after initial hashing. This comprehensive data allows for evaluating trade-offs between key size, signature size, signing speed, verification speed, and their scaling behavior with message payload.

\begin{table*}[!htbp]
\centering
\caption{Detailed signature performance and size metrics versus message size (macOS)}
\label{tab:appendix_sig_detailed}
\footnotesize
\resizebox{\linewidth}{!}{%
\begin{tabular}{@{}llllrrrrrrrrrrrrrr@{}}
\toprule
Alg.&Family&Type&Sec.&Key‑gen&Pub.&Sig‑1K&Sig‑10K&Sig‑100K&Sig‑1M&Sign‑1K&Sign‑10K&Sign‑100K&Sign‑1M&Ver‑1K&Ver‑10K&Ver‑100K&Ver‑1M\\
\midrule
Falcon‑512&Falcon&PQ&1&3.28&897&655&658&655&656&0.11&0.12&0.22&1.23&0.02&0.03&0.12&1.08\\
Falcon‑1024&Falcon&PQ&5&10.31&1\,793&1\,273&1\,272&1\,272&1\,274&0.21&0.22&0.32&1.27&0.04&0.05&0.14&1.09\\
Falcon‑pad‑512&Falcon‑pad&PQ&1&3.32&897&666&666&666&666&0.11&0.12&0.24&1.44&0.02&0.03&0.12&1.08\\
Falcon‑pad‑1024&Falcon‑pad&PQ&5&10.20&1\,793&1\,280&1\,280&1\,280&1\,280&0.21&0.22&0.32&1.27&0.04&0.05&0.14&1.09\\
SPHINCS+‑S2‑128f&Hash&PQ&1&0.64&32&17\,088&17\,088&17\,088&17\,088&15.05&15.10&15.45&18.56&0.91&0.93&1.09&2.70\\
SPHINCS+‑S2‑128s&Hash&PQ&1&41.09&32&7\,856&7\,856&7\,856&7\,856&310.81&310.37&310.99&312.37&0.32&0.33&0.50&2.11\\
SPHINCS+‑S2‑192f&Hash&PQ&3&0.95&48&35\,664&35\,664&35\,664&35\,664&24.61&24.68&24.85&26.89&1.34&1.35&1.46&2.49\\
SPHINCS+‑S2‑192s&Hash&PQ&3&59.60&48&16\,224&16\,224&16\,224&16\,224&553.84&561.00&559.74&556.11&0.48&0.49&0.60&1.66\\
SPHINCS+‑S2‑256f&Hash&PQ&5&2.45&64&49\,856&49\,856&49\,856&49\,856&50.66&50.50&50.69&52.71&1.36&1.37&1.47&2.52\\
SPHINCS+‑S2‑256s&Hash&PQ&5&39.47&64&29\,792&29\,792&29\,792&29\,792&489.49&489.74&492.98&494.28&0.68&0.71&0.81&1.88\\
SPHINCS+‑SK‑128f&Hash&PQ&1&0.96&32&17\,088&17\,088&17\,088&17\,088&21.04&21.21&21.30&23.16&1.26&1.28&1.36&2.33\\
SPHINCS+‑SK‑128s&Hash&PQ&1&57.65&32&7\,856&7\,856&7\,856&7\,856&438.88&445.07&442.28&446.72&0.44&0.47&0.56&1.57\\
SPHINCS+‑SK‑192f&Hash&PQ&3&1.38&48&35\,664&35\,664&35\,664&35\,664&35.87&36.06&36.51&39.40&1.96&1.95&2.07&3.20\\
SPHINCS+‑SK‑192s&Hash&PQ&3&88.63&48&16\,224&16\,224&16\,224&16\,224&800.78&798.61&799.04&800.76&0.68&0.69&0.79&1.77\\
SPHINCS+‑SK‑256f&Hash&PQ&5&3.68&64&49\,856&49\,856&49\,856&49\,856&73.51&73.70&73.86&75.82&1.96&1.97&2.07&3.05\\
SPHINCS+‑SK‑256s&Hash&PQ&5&58.42&64&29\,792&29\,792&29\,792&29\,792&698.14&698.43&699.10&702.15&0.98&0.98&1.08&2.09\\
ML‑DSA‑44&ML‑DSA&PQ&2&0.03&1\,312&2\,420&2\,420&2\,420&2\,420&0.16&0.16&0.26&1.21&0.04&0.05&0.14&1.10\\
ML‑DSA‑65&ML‑DSA&PQ&3&0.07&1\,952&3\,309&3\,309&3\,309&3\,309&0.26&0.27&0.36&1.31&0.06&0.07&0.17&1.12\\
ML‑DSA‑87&ML‑DSA&PQ&5&0.10&2\,592&4\,627&4\,627&4\,627&4\,627&0.31&0.31&0.41&1.41&0.10&0.11&0.20&1.21\\
MAYO‑1&Multivar.&PQ&1&0.05&1\,168&321&321&321&321&0.17&0.18&0.27&1.23&0.04&0.05&0.14&1.10\\
MAYO‑2&Multivar.&PQ&1&0.15&5\,488&180&180&180&180&0.24&0.25&0.34&1.28&0.03&0.04&0.13&1.08\\
MAYO‑3&Multivar.&PQ&3&0.20&2\,656&577&577&577&577&0.60&0.60&0.70&1.64&0.12&0.13&0.23&1.17\\
MAYO‑5&Multivar.&PQ&5&0.57&5\,008&838&838&838&838&1.51&1.52&1.61&2.56&0.28&0.29&0.38&1.33\\
c‑rsdp‑128‑bal&RSDP&PQ&1&0.01&77&12\,912&12\,912&12\,912&12\,912&0.63&0.63&0.71&1.50&0.36&0.36&0.44&1.23\\
c‑rsdp‑128‑fast&RSDP&PQ&1&0.01&77&19\,152&19\,152&19\,152&19\,152&0.35&0.35&0.44&1.22&0.20&0.20&0.29&1.06\\
c‑rsdp‑128‑small&RSDP&PQ&1&0.01&77&10\,080&10\,080&10\,080&10\,080&2.32&2.34&2.40&3.21&1.32&1.34&1.41&2.20\\
c‑rsdp‑192‑bal&RSDP&PQ&3&0.02&115&28\,222&28\,222&28\,222&28\,222&1.41&1.40&1.50&2.47&0.76&0.76&0.86&1.83\\
c‑rsdp‑192‑fast&RSDP&PQ&3&0.02&115&42\,682&42\,682&42\,682&42\,682&0.78&0.79&0.91&1.84&0.44&0.45&0.56&1.51\\
c‑rsdp‑192‑small&RSDP&PQ&3&0.02&115&23\,642&23\,642&23\,642&23\,642&3.29&3.29&3.38&4.36&1.73&1.74&1.83&2.81\\
c‑rsdp‑256‑bal&RSDP&PQ&5&0.04&153&51\,056&51\,056&51\,056&51\,056&2.47&2.48&2.59&3.57&1.24&1.25&1.36&2.33\\
c‑rsdp‑256‑fast&RSDP&PQ&5&0.04&153&76\,298&76\,298&76\,298&76\,298&1.52&1.51&1.62&2.59&0.86&0.86&0.96&1.94\\
c‑rsdp‑256‑small&RSDP&PQ&5&0.04&153&43\,592&43\,592&43\,592&43\,592&4.64&4.78&4.71&5.66&2.26&2.33&2.35&3.31\\
c‑rsdpg‑128‑bal&RSDP‑G&PQ&1&0.00&54&9\,236&9\,236&9\,236&9\,236&0.45&0.45&0.53&1.34&0.26&0.26&0.34&1.14\\
c‑rsdpg‑128‑fast&RSDP‑G&PQ&1&0.00&54&12\,472&12\,472&12\,472&12\,472&0.23&0.24&0.31&1.10&0.13&0.14&0.22&1.01\\
c‑rsdpg‑128‑small&RSDP‑G&PQ&1&0.00&54&7\,956&7\,956&7\,956&7\,956&1.58&1.58&1.65&2.43&0.90&0.91&0.98&1.76\\
c‑rsdpg‑192‑bal&RSDP‑G&PQ&3&0.01&83&23\,380&23\,380&23\,380&23\,380&0.60&0.61&0.70&1.65&0.36&0.38&0.47&1.42\\
c‑rsdpg‑192‑fast&RSDP‑G&PQ&3&0.01&83&27\,404&27\,404&27\,404&27\,404&0.46&0.47&0.57&1.51&0.28&0.29&0.39&1.34\\
c‑rsdpg‑192‑small&RSDP‑G&PQ&3&0.01&83&18\,188&18\,188&18\,188&18\,188&2.17&2.17&2.26&3.22&1.37&1.38&1.47&2.42\\
c‑rsdpg‑256‑bal&RSDP‑G&PQ&5&0.01&106&40\,134&40\,134&40\,134&40\,134&1.01&1.03&1.11&2.07&0.61&0.62&0.71&1.68\\
c‑rsdpg‑256‑fast&RSDP‑G&PQ&5&0.01&106&48\,938&48\,938&48\,938&48\,938&0.77&0.78&0.87&1.85&0.48&0.49&0.58&1.55\\
c‑rsdpg‑256‑small&RSDP‑G&PQ&5&0.01&106&32\,742&32\,742&32\,742&32\,742&2.76&2.78&2.86&3.81&1.65&1.67&1.76&2.72\\
ECDSA‑256&ECDSA&C&1&0.01&91&70&70&72&70&0.06&0.03&0.06&0.34&0.05&0.06&0.08&0.36\\
ECDSA‑384&ECDSA&C&3&0.20&120&103&103&103&102&0.30&0.31&0.34&0.63&0.59&0.60&0.63&0.91\\
ECDSA‑521&ECDSA&C&5&0.22&158&138&139&138&138&0.36&0.36&0.40&0.71&0.61&0.61&0.66&0.96\\
Ed25519&EdDSA&C&0&0.07&44&64&64&64&64&0.07&0.08&0.18&1.21&0.15&0.16&0.21&0.71\\
RSA‑2048&RSA&C&1&54.97&294&256&256&256&256&1.79&1.79&1.83&2.10&0.06&0.06&0.09&0.39\\
RSA‑3072&RSA&C&3&162.76&422&384&384&384&384&3.88&3.89&3.92&4.23&0.10&0.11&0.13&0.43\\
RSA‑4096&RSA&C&5&411.65&550&512&512&512&512&7.00&7.09&7.16&7.45&0.16&0.17&0.20&0.50\\
\bottomrule
\end{tabular}}%
\end{table*}

\subsection{Platform Performance Ratios}

To quantify the performance differences attributable to hardware and potentially underlying software library optimizations between platforms, we calculated performance ratios relative to the macOS reference platform (macOS M4 ARM). A ratio greater than 1 indicates that the comparison platform (Ubuntu Intel x86 or Raspberry Pi 4 ARM Cortex-A72) is slower than macOS for that specific operation and algorithm family. Conversely, a ratio less than 1 indicates the comparison platform was faster. These ratios help isolate the impact of the platform itself from the inherent cost of the cryptographic algorithm.

\subsubsection{KEM Performance Ratios}
\begin{figure}[!htbp]
  \centering
  \begin{subfigure}{0.48\linewidth}
    \includegraphics[width=\linewidth,
                     trim=5pt 5pt 25pt 15pt,clip]
                     {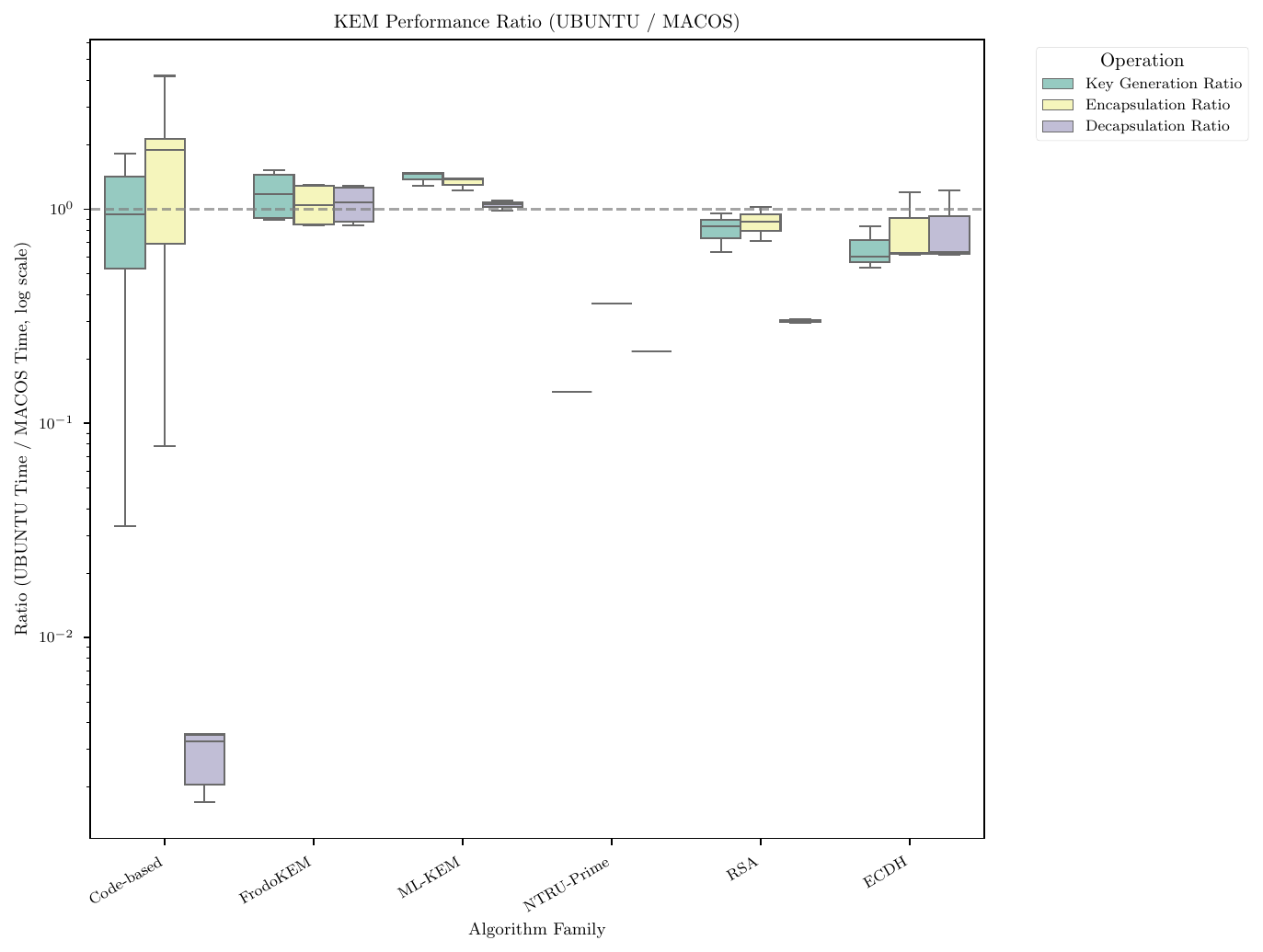}
    \caption{Ubuntu / macOS}
    \label{fig:kem_ratio_ubuntu_app}
  \end{subfigure}\hfill
  \begin{subfigure}{0.48\linewidth}
    \includegraphics[width=\linewidth,
                     trim=5pt 5pt 25pt 15pt,clip]
                     {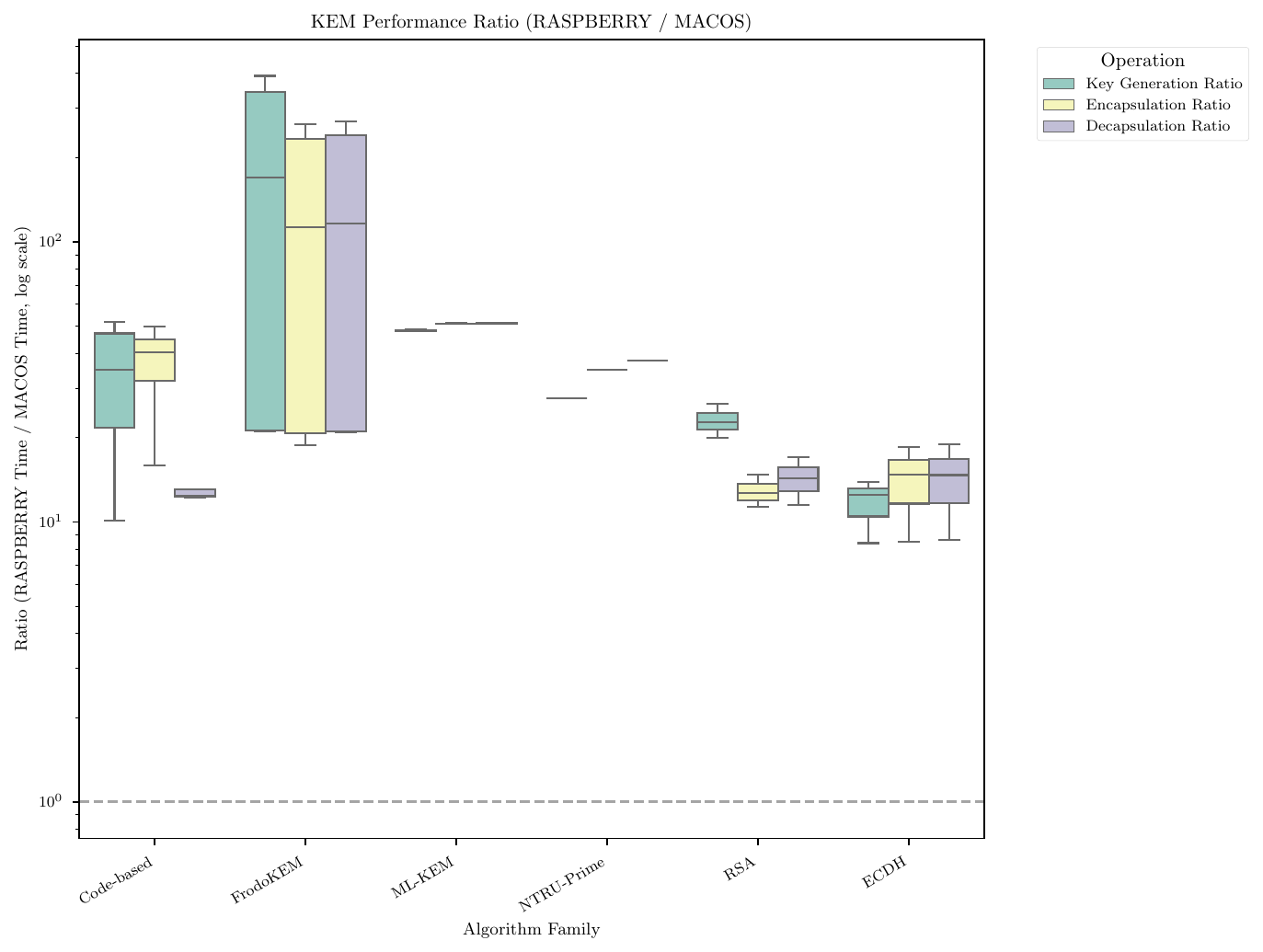}
    \caption{Raspberry Pi / macOS}
    \label{fig:kem_ratio_raspberry_app}
  \end{subfigure}
  \caption{KEM performance ratios relative to the macOS reference platform
  (log scale).  Ratios $>1$ mean the comparison platform is slower.}
  \label{fig:kem_ratios_app}
\end{figure}

Figures~\ref{fig:kem_ratio_ubuntu_app} and~\ref{fig:kem_ratio_raspberry_app} visually represent the distribution of these performance ratios using box plots for KEM operations (Key Generation, Encapsulation, Decapsulation). Figure~\ref{fig:kem_ratio_ubuntu_app} compares the Ubuntu desktop performance to macOS, while Figure~\ref{fig:kem_ratio_raspberry_app} compares the resource-constrained Raspberry Pi to macOS. The plots are grouped by algorithm family to reveal potential family-specific sensitivities to platform changes. The logarithmic scale emphasizes the magnitude of the slowdowns observed, particularly on the Raspberry Pi.

Table~\ref{tab:appendix_kem_platform_ratios} provides a statistical summary (mean, median, min, max) of these performance ratios for each algorithm family and operation, complementing the visual representation in the figures. This allows for a more quantitative assessment of the average, typical, and extreme performance differences observed between the platforms for each family.

\begin{table*}[!htbp]
  \centering
  \caption{Summary of KEM performance ratios relative to macOS}
  \label{tab:appendix_kem_platform_ratios}
  \footnotesize
  \resizebox{0.9\linewidth}{!}{%
  \begin{tabular}{@{}cll*{12}{r}@{}}
    \toprule
    &&&
    \multicolumn{4}{c}{Key‑generation} &
    \multicolumn{4}{c}{Encapsulation}  &
    \multicolumn{4}{c@{}}{Decapsulation} \\
    \cmidrule(lr){4-7}\cmidrule(lr){8-11}\cmidrule(lr){12-15}
    Family & Type & Platform &
    mean & med & min & max &
    mean & med & min & max &
    mean & med & min & max \\
    \midrule
    Code‑based & PQ & Raspberry & 39.94 & 34.90 & 10.14 & 139.27 & 36.98 & 40.24 & 15.96 & 49.73 & 18.81 & 12.41 & 12.23 & 47.67 \\
    Code‑based & PQ & Ubuntu    &  1.23 &  0.95 &  0.03 &   4.34 &  1.75 &  1.89 &  0.08 &  4.37 &  0.50 &  0.00 &  0.00 &  4.33 \\
    ECDH       & C  & Raspberry & 11.60 & 12.53 &  8.40 &  13.88 & 13.92 & 14.77 &  8.49 & 18.50 & 14.10 & 14.70 &  8.61 & 18.97 \\
    ECDH       & C  & Ubuntu    &  0.66 &  0.60 &  0.53 &   0.84 &  0.81 &  0.62 &  0.61 &  1.20 &  0.82 &  0.63 &  0.61 &  1.22 \\
    FrodoKEM   & PQ & Raspberry & 187.74 & 169.85 &  21.04 & 391.65 & 128.71 & 112.88 & 18.82 & 264.22 & 132.35 & 116.22 & 20.88 & 269.18 \\
    FrodoKEM   & PQ & Ubuntu    &   1.19 &   1.17 &   0.90 &   1.52 &   1.06 &   1.05 &  0.84 &   1.30 &   1.07 &   1.08 &  0.84 &   1.28 \\
    ML‑KEM     & PQ & Raspberry & 48.25 & 48.20 & 47.96 & 48.61 & 51.09 & 50.98 & 50.88 & 51.41 & 51.12 & 51.09 & 50.86 & 51.39 \\
    ML‑KEM     & PQ & Ubuntu    &  1.41 &  1.47 &  1.28 &  1.48 &  1.33 &  1.38 &  1.22 &  1.39 &  1.05 &  1.06 &  0.99 &  1.10 \\
    NTRU‑Prime & PQ & Raspberry & 27.63 & 27.63 & 27.63 & 27.63 & 34.99 & 34.99 & 34.99 & 34.99 & 37.75 & 37.75 & 37.75 & 37.75 \\
    NTRU‑Prime & PQ & Ubuntu    &  0.14 &  0.14 &  0.14 &  0.14 &  0.36 &  0.36 &  0.36 &  0.36 &  0.22 &  0.22 &  0.22 &  0.22 \\
    RSA        & C  & Raspberry & 23.00 & 22.70 & 19.93 & 26.37 & 12.90 & 12.65 & 11.30 & 14.76 & 14.26 & 14.29 & 11.48 & 17.00 \\
    RSA        & C  & Ubuntu    &  0.81 &  0.83 &  0.63 &  0.96 &  0.87 &  0.88 &  0.71 &  1.02 &  0.30 &  0.30 &  0.29 &  0.31 \\
    \bottomrule
  \end{tabular}}%
\end{table*}

\subsubsection{Signature Performance Ratios}
\begin{figure}[!htbp]
  \centering
  \begin{subfigure}{0.48\linewidth}
    \includegraphics[width=\linewidth,
                     trim=5pt 5pt 25pt 15pt,clip]
                     {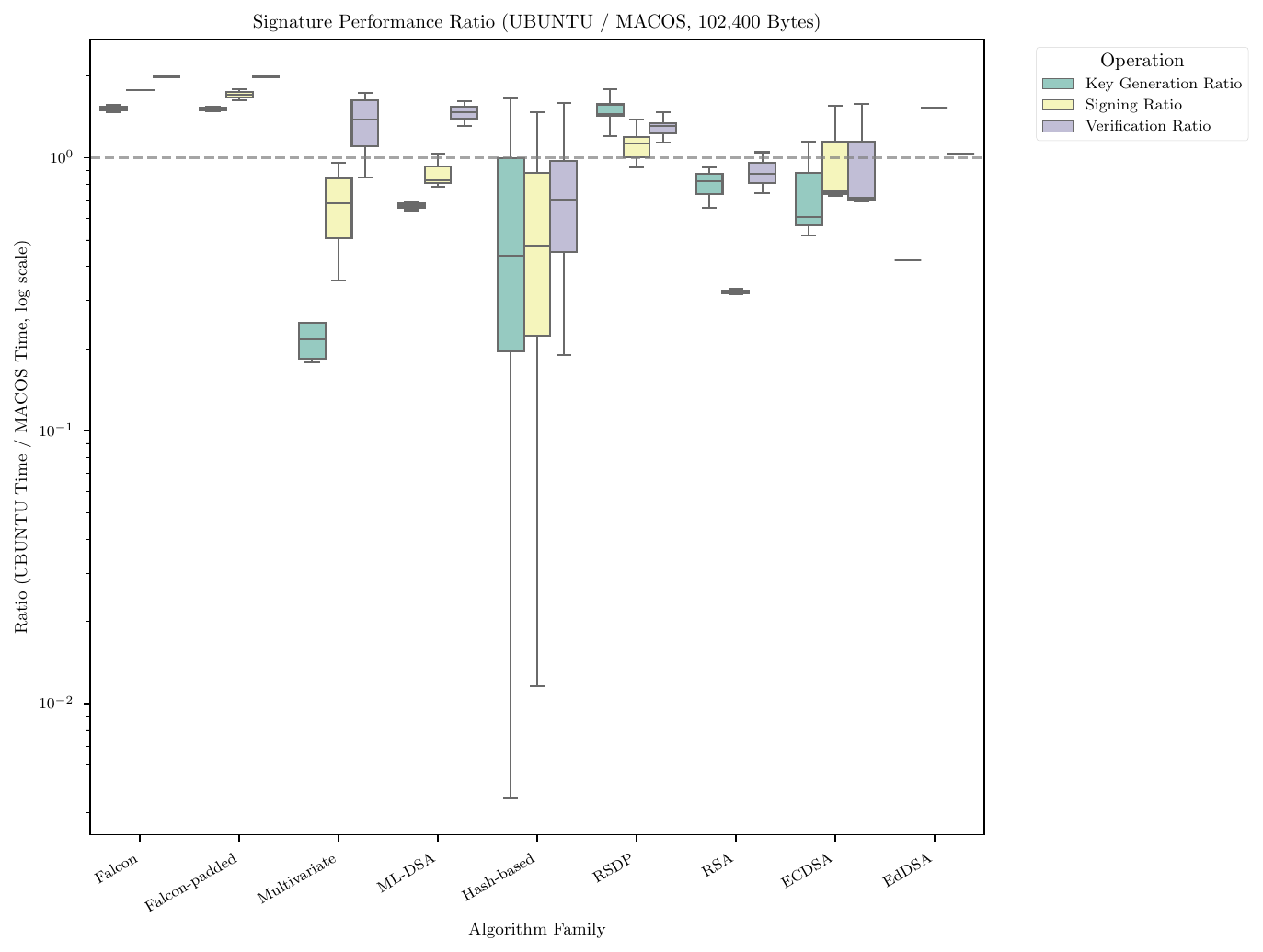}
    \caption{Ubuntu / macOS}
    \label{fig:sig_ratio_ubuntu_app}
  \end{subfigure}\hfill
  \begin{subfigure}{0.48\linewidth}
    \includegraphics[width=\linewidth,
                     trim=5pt 5pt 25pt 15pt,clip]
                     {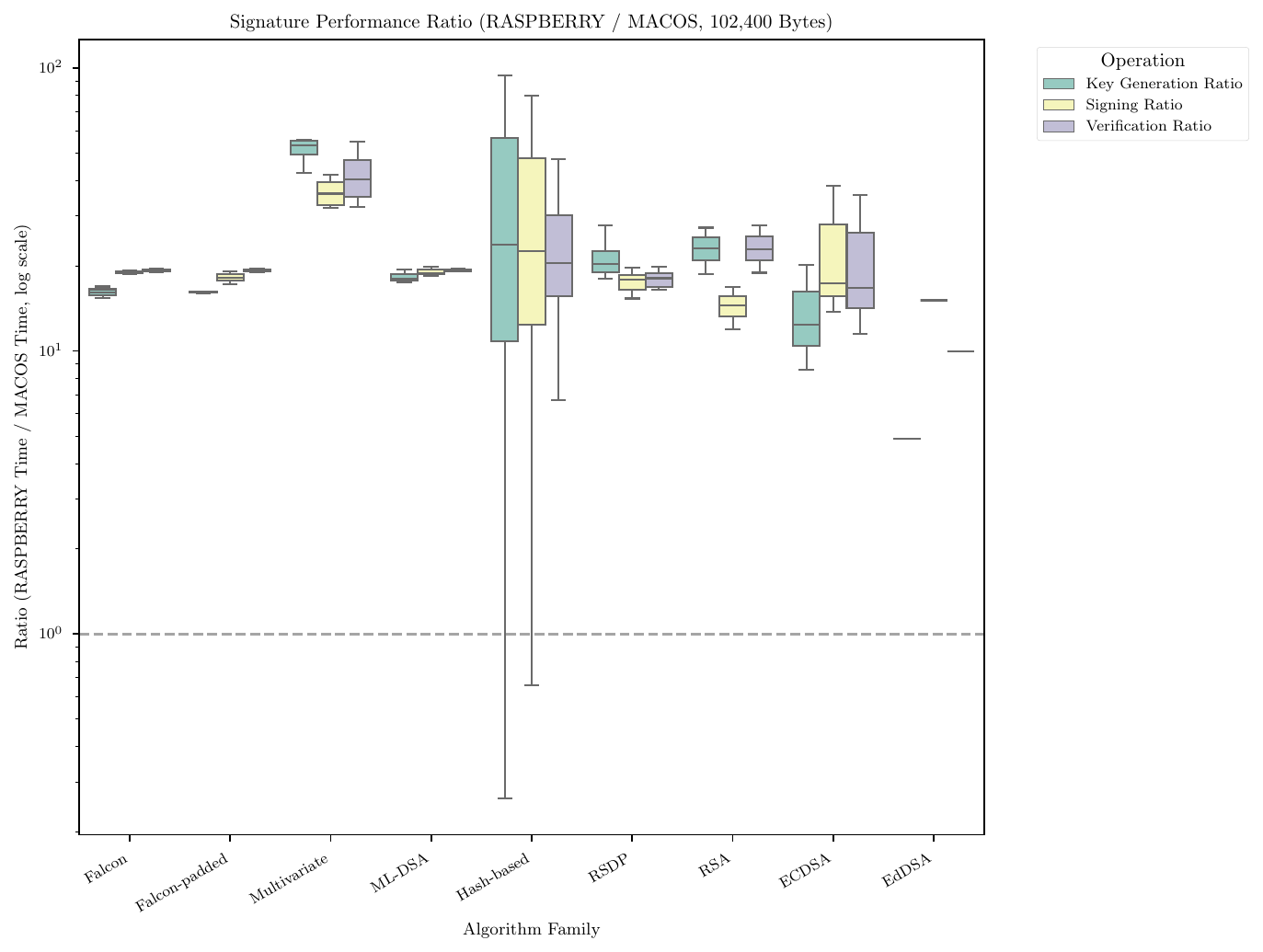}
    \caption{Raspberry Pi / macOS}
    \label{fig:sig_ratio_raspberry_app}
  \end{subfigure}
  \caption{Signature performance ratios relative to the macOS reference platform
  for a \SI{100}{\kilo\byte} message (log scale).  Ratios $>1$ mean the
  comparison platform is slower.}
  \label{fig:sig_ratios_app}
\end{figure}

Similarly, Figures~\ref{fig:sig_ratio_ubuntu_app} and~\ref{fig:sig_ratio_raspberry_app} illustrate the performance ratios for signature operations (Key Generation, Signing, Verification) relative to macOS. These plots use the \SI{100}{\kilo\byte} message size results as a representative example to show the platform impact on signing and verification for a moderately sized payload. Again, the ratios highlight the performance difference between Ubuntu/macOS and Raspberry Pi/macOS, grouped by signature algorithm family.

Table~\ref{tab:sig_platform_ratios_app} provides the corresponding statistical summary of these signature performance ratios for the \SI{100}{\kilo\byte} message case. This table quantifies the average, median, and range of slowdown factors observed on Ubuntu and Raspberry Pi compared to macOS for key generation, signing, and verification across different signature families. Note that while these ratios are shown for the \SI{100}{\kilo\byte} message size, similar analyses for other message sizes (available in the project repository) confirm the general trends, although the exact ratio values may differ due to the message size scaling behavior discussed in the main text.

\begin{table}[H]
  \centering
  \caption{Summary of signature performance ratios relative to macOS (100 kB message)}
  \label{tab:sig_platform_ratios_app}
  \footnotesize
  \resizebox{0.8\linewidth}{!}{%
    \begin{tabular}{@{}cllrrrrrrrrrrrr@{}}
      \toprule
      &&&\multicolumn{4}{c}{Key‑generation}
      &\multicolumn{4}{c}{Signing}
      &\multicolumn{4}{c@{}}{Verification}\\
      \cmidrule(lr){4-7}\cmidrule(lr){8-11}\cmidrule(lr){12-15}
      Family & Type & Platform
             & mean & med & min & max
             & mean & med & min & max
             & mean & med & min & max\\
      \midrule
      ECDSA & C  & Raspberry & 13.68 & 12.35 &  8.56 & 20.13 & 23.21 & 17.39 & 13.78 & 38.46 & 21.28 & 16.73 & 11.51 & 35.59\\
      ECDSA & C  & Ubuntu    &  0.76 &  0.61 &  0.52 &  1.15 &  1.01 &  0.75 &  0.73 &  1.55 &  0.99 &  0.71 &  0.69 &  1.57\\
      EdDSA & C  & Raspberry &  4.90 &  4.90 &  4.90 &  4.90 & 15.10 & 15.10 & 15.10 & 15.10 &  9.99 &  9.99 &  9.99 &  9.99\\
      EdDSA & C  & Ubuntu    &  0.42 &  0.42 &  0.42 &  0.42 &  1.52 &  1.52 &  1.52 &  1.52 &  1.04 &  1.04 &  1.04 &  1.04\\
      Falcon & PQ & Raspberry & 16.14 & 16.14 & 15.39 & 16.89 & 18.96 & 18.96 & 18.74 & 19.18 & 19.25 & 19.25 & 18.91 & 19.60\\
      Falcon & PQ & Ubuntu    &  1.52 &  1.52 &  1.47 &  1.56 &  1.77 &  1.77 &  1.76 &  1.77 &  1.98 &  1.98 &  1.97 &  1.99\\
      Falcon‑padded & PQ & Raspberry & 16.14 & 16.14 & 16.02 & 16.26 & 18.19 & 18.19 & 17.27 & 19.11 & 19.25 & 19.25 & 19.00 & 19.49\\
      Falcon‑padded & PQ & Ubuntu    &  1.51 &  1.51 &  1.48 &  1.53 &  1.70 &  1.70 &  1.63 &  1.78 &  1.98 &  1.98 &  1.97 &  2.00\\
      Hash‑based & PQ & Raspberry & 263.84 & 23.74 &  0.26 & 2275.70 & 112.98 & 22.49 &  0.66 & 864.97 & 24.71 & 20.40 &  6.72 & 67.98\\
      Hash‑based & PQ & Ubuntu    &  4.59 &  0.44 &  0.00 &  39.64 &  2.10 &  0.48 &  0.01 &  15.93 &  0.77 &  0.70 &  0.19 &  1.91\\
      ML‑DSA & PQ & Raspberry & 18.29 & 18.01 & 17.45 & 19.42 & 19.06 & 18.86 & 18.44 & 19.87 & 19.26 & 19.23 & 19.05 & 19.51\\
      ML‑DSA & PQ & Ubuntu    &  0.67 &  0.67 &  0.64 &  0.69 &  0.88 &  0.83 &  0.79 &  1.04 &  1.46 &  1.47 &  1.31 &  1.61\\
      Multivariate & PQ & Raspberry & 51.29 & 53.38 & 42.61 & 55.81 & 36.43 & 35.97 & 31.93 & 41.84 & 41.96 & 40.33 & 32.22 & 54.97\\
      Multivariate & PQ & Ubuntu    &  0.22 &  0.22 &  0.18 &  0.25 &  0.67 &  0.68 &  0.35 &  0.96 &  1.34 &  1.39 &  0.85 &  1.73\\
      RSA & C & Raspberry & 23.00 & 23.02 & 18.73 & 27.27 & 14.43 & 14.52 & 11.97 & 16.81 & 23.19 & 22.88 & 18.89 & 27.80\\
      RSA & C & Ubuntu    &  0.80 &  0.82 &  0.66 &  0.92 &  0.32 &  0.32 &  0.32 &  0.33 &  0.89 &  0.88 &  0.74 &  1.05\\
      RSDP & PQ & Raspberry & 21.23 & 20.30 & 17.94 & 27.78 & 17.54 & 17.83 & 15.33 & 19.75 & 17.91 & 18.07 & 16.48 & 19.89\\
      RSDP & PQ & Ubuntu    &  1.48 &  1.45 &  1.17 &  1.85 &  1.13 &  1.13 &  0.93 &  1.38 &  1.29 &  1.31 &  1.14 &  1.47\\
      \bottomrule
    \end{tabular}}%
\end{table}

\end{document}